\documentclass[
superscriptaddress,
reprint,
amsmath,amssymb,
aps,
pra
]{revtex4-2}
\usepackage[english]{babel}
\usepackage{makecell}
\usepackage[colorlinks=false]{hyperref}

\usepackage{tikz}

\usepackage{graphicx}
\usepackage{dcolumn}
\usepackage{bm}

\usepackage{hhline}
\usepackage{multirow}
\usepackage{dsfont}
\usepackage{physics}
\usepackage{pifont}
\usepackage{xcolor}

\newcommand{\id}{\mathbb I}

\begin{document}

\title{Bell-pair distillation under imperfect operations}
\author{Andr\'{a}s Kiny\'{o}}
\email{kinyeaux@student.elte.hu}
\affiliation{E\"otv\"os Lor\'and University, P.O.\ Box 32, H-1518, Hungary}
\affiliation{HUN-REN Wigner Research Centre for Physics, 1525 P.O. Box 49, Budapest, Hungary}
\author{Tam\'as Kiss}
\email{kiss.tamas@wigner.hu}
\affiliation{HUN-REN Wigner Research Centre for Physics, 1525 P.O. Box 49, Budapest, Hungary}
\author{Orsolya K\'alm\'an}
\email{kalman.orsolya@wigner.hu}
\affiliation{HUN-REN Wigner Research Centre for Physics, 1525 P.O. Box 49, Budapest, Hungary}

\begin{abstract}
We recently proposed an entanglement distillation protocol based on local operations and classical communication that probabilistically and universally distills a maximally entangled Bell pair from multiple copies of an unknown two-qubit state [Phys. Rev. Lett. 135, 260202 (2025)]. Here, we investigate the robustness of this protocol against imperfections in its implementation. We consider several incoherent and coherent gate-error models, as well as measurement readout errors, and determine the regimes in which distillation enhances the fidelity with the target Bell state. We find that the protocol remains effective over substantial error ranges, although its robustness depends strongly on the type of error. We further analyze asymmetric errors between the two parties and identify conditions under which complementary postselection outcomes yield identical distillation dynamics, allowing the success probability to be enhanced.
\end{abstract}

\maketitle

\section{Introduction}
Entanglement is a central resource in quantum technologies, underpinning a wide range of quantum computation and communication protocols~\cite{Bennett1993}. In particular, distributed Bell pairs serve as fundamental building blocks of many schemes as well as resources for multipartite entanglement. However, their reliable preparation in distributed settings remains challenging due to noise affecting both initialization and distribution.

Nevertheless, several experiments have demonstrated entanglement distribution between remote nodes on a variety of physical platforms, including color centers in diamond~\cite{Pompili2022,Stas2022,Knaut2024}, neutral atoms~\cite{vanLeent2022}, trapped ions~\cite{Stephenson2020}, and superconducting circuits~\cite{Leung2019,Magnard2020}, in some cases over tens of kilometers of deployed telecom fiber~\cite{Knaut2024,vanLeent2022}. Meanwhile, high-fidelity local operations have been achieved on these and other platforms~\cite{Bartling2025,Ademi2025,Steinacker2025,Evered2023,Li2024,Weber2024,Pingault2017}. Still, the fidelity of distributed Bell pairs remains limited by noise during preparation and transmission.

One approach to mitigating these imperfections is the use of LOCC (local operations and classical communication) distillation protocols \cite{Bennett_distill_1996,Bennett_distill_2_1996,Deutsch_distill_1996,Macchiavello_1998,Dur_Briegel_1998,Dur_Briegel_1999,Hsieh_2004,Torres_2016,Rozpedek_2018,Bernad_2022}. 
Quantum circuits that incorporate measurement and post-selection induce an effectively nonlinear evolution of quantum states when applied iteratively. With appropriate design, such circuits can drive a broad class of input states to converge toward a desired target state with a favorable error suppression \cite{Marty2026,Kalman2025}. This mechanism can be leveraged for entanglement distillation, progressively enhancing both the purity and entanglement of the states.

A recent example is the protocol proposed by Kálmán et al.~\cite{Kalman2025}, which aims to distill a Bell pair shared between two parties in a few iterations without any a priori information about the input state.
In~\cite{Kalman2025}, we showed that the protocol transforms any entangled pure input state, apart from a set of measure zero, into $\ket{\Phi_+}$ within two steps. Thus, it fully removes coherent preparation errors. It was recently proven to do so with optimal success probability among universal conclusive protocols built from concatenated two-qubit operations~\cite{Orthey_2026}. The term universality is used in different senses in the recent literature. Zang \textit{et al.} proved that no LOCC protocol acting on a finite number of copies can guarantee that the output fidelity never falls below the input fidelity for all entangled two-qubit states~\cite{Zang2025}. In the asymptotic regime, the distillable entanglement of an unknown state can nevertheless be reached~\cite{Tirone2026}. Our protocol is universal in the conclusive sense: whenever it succeeds, it outputs the target state without prior knowledge of the input.
For input states with moderate preparation noise, the protocol suppresses the initial noise quadratically after three iterations.
These results hold under the assumption of ideal quantum operations, i.e., in the absence of gate noise.

It has been known since the early proposals that imperfect local operations limit both the input fidelity required for purification and the output fidelity that can be attained~\cite{Dur_Briegel_1998,Dur_Briegel_1999,Giedke1999,DurBriegel2007}. The corresponding error thresholds have been determined for several protocols and noise models~\cite{Rohde2007,Zwerger2013}. Protocols have also been optimized for practical conditions~\cite{Rozpedek_2018,Krastanov2019,Jansen2022,Bernad_2022,PretiBernad2024}, including robustness against noisy local operations~\cite{FujiiYamamoto2009,Krastanov2019}. More recently, distillation was analyzed under realistic device noise, including damping, readout and coherent errors, and tested on superconducting hardware~\cite{Siddhu2025}. Noisy measurements were also shown to restrict distillation, which can be counteracted by purifying the measurements themselves~\cite{KimYunBae2025,Kim2026}. Nevertheless, experimental demonstrations of entanglement purification with photons~\cite{Pan2003,Hu2021,Ecker2021}, trapped ions~\cite{Reichle2006}, color centers in diamond~\cite{Kalb2017}, and superconducting circuits~\cite{Yan2022} show that distillation can enhance the fidelity even under realistic noise.

In this work, we investigate the robustness of the scheme of Kálmán et al.~\cite{Kalman2025} against imperfect operations. We consider incoherent gate errors as well as coherent gate errors, arising, e.g., from miscalibrated driving, which, unlike coherent preparation errors, are not automatically removed by the protocol. We also include readout errors, which are particularly relevant since the protocol relies on post-selection. Assessing the performance of the protocol under such imperfections is essential before its practical implementation on any given physical architecture.

The paper is organized as follows. We start by establishing the mathematical model to be used in the calculations (Section \ref{sec:2}.). Then we consider incoherent noise models (Sec.~\ref{sec:3}), comparing the feasibility of the protocol in the presence of each, and also provide estimates on how robust the scheme was if we implemented it on certain existing physical platforms. In Sec.~\ref{sec:4} our main focus is on coherent gate error, considering two different models. The effect of readout (measurement) errors is studied in Sec.~\ref{sec:5}, employing both balanced and biased errors. 
In Sec.~\ref{sec:6} we give some examples considering different error models simultaneously. Finally, in Sec.~\ref{sec:7} we discuss the probabilistic nature of the distillation scheme and propose an improvement for the state generation rate.

\section{The protocol}
\label{sec:2}

\subsection{No gate errors}

\begin{figure}[h]
\centering
\includegraphics[width=\linewidth]{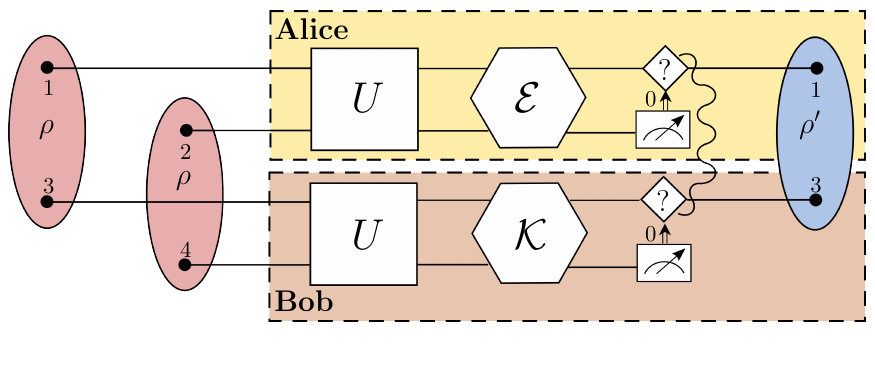}
\vspace{-6ex}
\caption{Circuit corresponding to one iteration of the protocol. In the presence of gate noise, noise channels $\mathcal E$ and $\mathcal K$ act on qubits $(1,2)$ and $(3,4)$, respectively. (Figure adapted from~\cite{Kalman2025}.)}
\label{fig:1_pure_protocol}
\end{figure}

A single iteration of the protocol~\cite{Kalman2025} consists of three main steps. First, Alice and Bob distribute their qubits: Initially, Alice prepares two entangled pairs—labelled (1,3) and (2,4)—in the same (possibly mixed) state. She then sends qubits 3 and 4 to Bob. Both parties subsequently apply the same unitary operation to their respective qubit pairs. Next, they measure qubits 2 and 4 (in the computational basis) and communicate the outcomes via a classical channel. If both measurements yield 0, they keep the remaining qubits (1 and 3), which are then in a transformed entangled state (see Fig.~\ref{fig:1_pure_protocol}). By repeating this procedure once more, the parties obtain two pairs of distributed qubits in the same state, which can be used as input for the next iteration. (Note that every further iteration of the protocol effectively doubles the number of necessary input qubit pairs.)

The unitary transformation applied by Alice and Bob is
\begin{equation}
U_{\pm} = \frac{1}{\sqrt{2}}
\begin{bmatrix}
0 & 1 & \pm1 & 0\\
1 & 0 & 0 & \pm1\\
0 & 1 & \mp1 & 0\\
1 & 0 & 0 & \mp1
\end{bmatrix}.
\end{equation}

The two unitaries $U_+$ and $U_-$ generate different dynamical evolutions, which are structurally similar. In what follows, we restrict our attention to $U_+$ (hereafter denoted simply by $U$), noting that all results can be obtained analogously for $U_-$. Note that $U$ can easily be decomposed into elementary quantum gates:
\begin{equation}
\label{eq:U}
U = (H \otimes \id)\, U_{\mathrm{CNOT}}\, (\id \otimes X),
\end{equation}
where $\id$ is the two-dimensional identity matrix.

We now turn to a mathematical description of a single iteration step. When Alice prepares and distributes the qubits, the global state is given by
\begin{equation}
\Omega = S_{23}\,(\rho \otimes \rho)\, S_{23},
\end{equation}
where $S_{23} = \id \otimes S \otimes \id$ exchanges qubits 2 and 3. This state is then processed by the two local quantum computers, where both parties apply the unitary $U$:
\begin{equation}
\label{eq:Omegaprime}
\Omega \mapsto \Omega' = (U \otimes U)\, \Omega \, (U \otimes U)^{\dagger}.
\end{equation}

Finally, measuring qubits 2 and 4 and post-selecting on the outcome '$00$' can be described by projecting onto $\ket{0}$ on these subsystems and tracing them out:
\begin{equation}
\label{eq:rhoprime_pure_protocol}
\rho' = \mathrm{Tr}_{2,4}(P_0\, \Omega' \, P_0), \quad
P_0 = \id \otimes \ket{0}\bra{0} \otimes \id \otimes \ket{0}\bra{0}.
\end{equation}

Projection onto the above-mentioned subspace of the four-qubit Hilbert space, followed by tracing out two subsystems, constitutes a non–trace-preserving transformation. Consequently, the resulting density matrix satisfies $\Tr\rho' \leq 1$. This trace can be interpreted as the \textit{success probability}, i.e., the probability that the measurement outcome '$00$' is obtained on qubits 2 and 4 in a given iteration. Conditioning on this event, it is convenient to renormalize the state according to $\rho' \mapsto \rho'/\Tr\rho'$.

\subsection{Introducing gate errors}

Quantum gate errors can be modeled using the quantum channel formalism. Quantum channels are convex linear, trace-preserving, completely positive (CPTP) maps, and are most conveniently described via their operator-sum (Kraus operator) representation~\cite{NielsenChuang2010}: 

\begin{equation}
    \mathcal{E}(\rho) = \sum_k E_k\,\rho\, E_k^{\dagger}, \qquad \sum_k E_k^{\dagger} \, E_k = \id.
\end{equation}

In order to incorporate quantum gate errors into the protocol, we insert such $\mathcal{E}$ channels after each unitary gate $U$, acting independently on each subsystem; a schematic of the corresponding circuit is shown in Fig.~\ref{fig:1_pure_protocol}. The transformation of the input state $\Omega$ is therefore modified as
\begin{equation}
    \label{eq:Omegaprime_general_gatenoise}
    \Omega' = [\mathcal E \otimes \mathcal K] \left([U\otimes U]\, \Omega [U\otimes U]^{\dagger}\right).
\end{equation}

This model enables the description of gate noise acting independently on the two local quantum processors, which is physically well motivated, as there is no guarantee that the two devices are identical. The tensor product of quantum channels can be naturally expressed in the Kraus operator representation as
\begin{equation}
    [\mathcal{E}\otimes \mathcal{K}](\rho_{AB}) = \sum_{k,l}(E_k\otimes K_l) \, \rho_{AB} \, (E_k\otimes K_l)^{\dagger},
\end{equation}
where $\mathcal{E}(\rho_A) = \sum_kE_k\,\rho_A\,E_k^{\dagger}$ and $ \mathcal{K}(\rho_B) = \sum_l K_l \, \rho_B \, K_l^{\dagger}$ are quantum channels acting on the two subsystems.

Combining Eqs.~\eqref{eq:rhoprime_pure_protocol} and \eqref{eq:Omegaprime_general_gatenoise}, the resulting state after one iteration can be written as
\begin{equation}
    \rho' = \Tr_{2,4}(P_0 \cdot[\mathcal E\otimes\mathcal K](\Omega')\cdot P_0^{\dagger}).
\end{equation}
The resulting density matrix is, in general, subnormalized, with $\Tr \rho' \leq 1$, reflecting the probabilistic nature of the post-selection. As in the noiseless case, this trace corresponds to the success probability of the protocol, and the state must be renormalized accordingly.

\section{Analysis of incoherent gate errors}
\label{sec:3}

With the mathematical framework established above, we now analyze the feasibility of the protocol in the presence of different types of gate errors. Since our primary interest lies in the effect of noise introduced in each iteration, we restrict our attention to input states of the form
\begin{equation}
\label{eq:rho0}
    \rho_0(\varepsilon) = (1-\varepsilon)\cdot \ket{\Phi_+}\bra{\Phi_+} + \varepsilon\cdot \dfrac{\id_4}{4},
\end{equation}
where $\id_4$ denotes the four-dimensional identity matrix.
Such states are commonly used as a model of preparation white noise, where $\varepsilon$ quantifies the noise strength, i.e., the deviation from a pure Bell state. As we will see later, some features of the interplay between different sources of infidelity cannot be captured using only this type of input state, however, one can cover many cases under such assumptions.

In the following, we consider several standard incoherent quantum noise channels: depolarizing noise, amplitude damping, bit-flip errors, and phase-flip (dephasing) errors. Our main objective is to demonstrate that the protocol remains robust in the presence of these incoherent error mechanisms. To keep the analysis tractable and to allow a detailed exploration of the relevant parameter regimes, we restrict ourselves to simplified channel models with only a small number of parameters. For a specific physical architecture, these models may be replaced by platform-dependent noise channels and examined over a more narrowly defined region of parameter space. Such architecture-specific examples lie outside the scope of this work; here, our aim is to provide a broader overview of the protocol’s behavior under representative forms of noise.

\subsection{Depolarization error}

One of the simplest gate noise model is the depolarizing channel, which acts on a two-qubit state as
\begin{equation}
    \mathcal{D}^{(\lambda)}(\rho) = (1-\lambda)\,\rho + \lambda\, \dfrac{\id_4}{4}.
\end{equation}
This channel admits a straightforward interpretation: With probability $1-\lambda$ the state is left unchanged, while with probability $\lambda$ it is replaced by the maximally mixed state. A corresponding set of Kraus operators is given by
\begin{equation}
    \begin{aligned}
    D_{00}^{(\lambda)} &= \sqrt{1-\dfrac{15\lambda}{16}}\, \id, \quad \\ 
    D_{\mu\nu}^{(\lambda)} &= \sqrt{\dfrac{\lambda}{16}}\, \sigma_{\mu}\otimes\sigma_{\nu},\quad (\mu,\nu)\neq(0,0),
\end{aligned}
\end{equation}
with $ \sigma_{\mu},\sigma_{\nu}\in\{I,X,Y,Z\}$. Note that this decomposition is not unique. In order to account for (possibly different) noise on Alice's and Bob's sides, we consider the channel $\mathcal D^{(\lambda_A)} \otimes \mathcal{D}^{(\lambda_B)}$.

The performance of a given distillation procedure can be assessed, for instance, by calculating the fidelity of the output state with respect to the $\ket{\Phi_+}$ Bell state:
\begin{equation}
\label{eq:fidelity}
\mathcal{F}(\rho) = \bra{\Phi_+}\rho\ket{\Phi_+} =\dfrac{1}{2} \left(\rho_{11} + \rho_{14} + \rho_{41} + \rho_{44}\right),
\end{equation}
where the matrix elements of $\rho$ are taken in the computational basis. We begin by investigating how the fidelity evolves throughout the iterations of the protocol, starting from a fixed input state $\rho_0(\varepsilon)$, for different values of $\lambda=\lambda_A=\lambda_B$. 

\begin{figure}
\centering
\includegraphics[width=\linewidth]{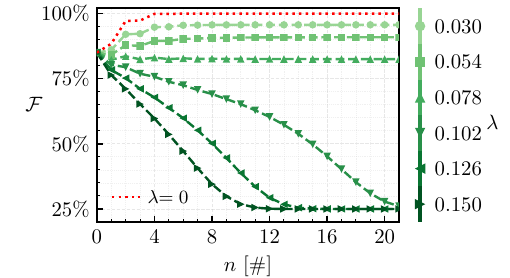}
\caption{Fidelity of the output state with respect to $\ket{\Phi_+}$ after successive protocol iterations for the input state  $\rho_0(\varepsilon=0.2)$. Each curve corresponds to a different value of the depolarization parameter $\lambda$. A red dotted line shows the noiseless case ($\lambda=0$).}
\label{fig:2_depolarization_steps}
\end{figure}

The results are shown in Fig.~\ref{fig:2_depolarization_steps} for an input state containing a relatively high level of initial noise ($\varepsilon=0.2$). It is immediately apparent that the protocol remains robust against moderate depolarization noise, revealing an interesting interplay between preparation noise and gate noise: When $\lambda$ is sufficiently small compared to $\varepsilon$, the state converges toward an output with high fidelity. However, for larger values of $\lambda$, the protocol instead drives the state toward the maximally mixed state (whose fidelity equals $0.25$). Another important feature to note is that if the depolarization noise is low enough, so that the output fidelity is larger than the input fidelity, the distillation (convergence to a stable output) is already achieved after four iterations (see cases with $\lambda< 0.05$).

The next natural step is to examine the interplay between $\varepsilon$ and $\lambda$ more closely and to identify a critical value of the latter, above which the protocol gets inefficient. To this end, we sample $\varepsilon$ and $\lambda$ on a sufficiently dense grid and calculate the output fidelity after four iterations, motivated by the above-mentioned observation that distillation is practically achieved within four iterations. For low noise levels, i.e., a high input fidelity, application of the protocol may even be disadvantageous if $\lambda$ is so large that it results in an output fidelity that is lower than the initial one. In order to account for this, we calculate the \textit{fidelity enhancement} 
\begin{equation}
\delta\mathcal F= \mathcal F(\rho_{\rm out}) - \mathcal{F}(\rho_{\rm in}),
\end{equation}
and define the critical value $\lambda_c(\varepsilon)$ by the condition $\delta \mathcal F=0$.

\begin{figure}
\centering
\begin{tikzpicture}
        \node[anchor=south west, inner sep=0] (img) {
            \includegraphics[width=\linewidth]
            {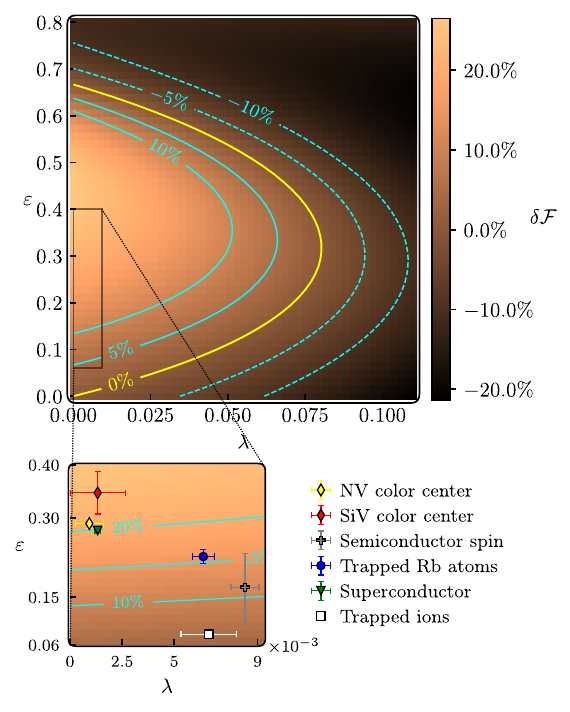}
        };

        \begin{scope}[x={(img.south east)}, y={(img.north west)}]
            \node[anchor=center] at (0.875,0.31){\scriptsize \cite{Pompili2022,Bartling2025}};
            \node[anchor=center] at (0.88,0.274){\scriptsize \cite{Stas2022,Knaut2024}};
            \node[anchor=center] at (0.938,0.239){\scriptsize \cite{Ademi2025, Steinacker2025}};
            \node[anchor=center] at (0.928,0.204){\scriptsize\cite{Evered2023, vanLeent2022}};
            \node[anchor=center] at (0.878,0.168){\scriptsize \cite{Li2024, Leung2019}};
            \node[anchor=center] at (0.845,0.134){\scriptsize \cite{Weber2024, Stephenson2020}};
        \end{scope}
    \end{tikzpicture}
\caption{ Change in the fidelity ($\delta\mathcal{F}$) after four iterations of the protocol, starting from input states with different $\varepsilon$ and including depolarization noise of varying strength ($\mathcal D^{(\lambda)}\otimes \mathcal D^{(\lambda)}$). The critical depolarization parameter curve, $\lambda_c(\varepsilon)$, corresponding to $\delta\mathcal F=0$, is highlighted with a yellow curve. The inset shows the parameter region corresponding to realistic devices as well as some specific examples (see references in the legend). Note that axes of the inset have different aspect ratios compared to those of the main figure. The effective error parameters were estimated as detailed in App. \ref{app:A}.}
\label{fig:3_deltaF_depolarization}
\end{figure}

The corresponding results are shown in Fig.~\ref{fig:3_deltaF_depolarization}. It can be seen that, in general, the protocol is beneficial when the initial noise is relatively large while the depolarization error remains moderate. Within the parameter range considered, the fidelity gain reaches a maximum of approximately $0.25$. We denoted the parameter range which is realistic with today's NISQ (noisy intermediate-scale quantum) devices by a thin black box, and in an inset plot we marked up some concrete physical platforms. It is clearly seen on which of these it could be beneficial to employ the protocol. Note that the aspect ratio of the inset differs from that of the main plot. The parameter values were estimated from reported state-preparation and gate infidelities, as described in App.~\ref{app:A}. In some cases, the two infidelities were extracted from different papers; therefore, these data should be regarded as order-of-magnitude reference points rather than reliable quantitative estimates.

\begin{figure}
\centering
\includegraphics[width=\linewidth]{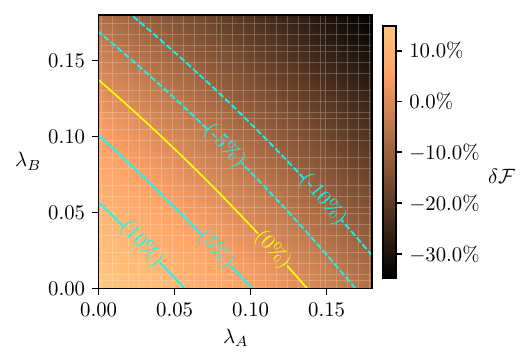}
\caption{Change in the fidelity ($\delta \mathcal F$) after four iterations for a fixed input state with white-noise parameter $\varepsilon=0.2$. An asymmetric depolarization channel was employed with parameters $\lambda_A$ and $\lambda_B$. The yellow curve corresponds to $\delta\mathcal F=0$.}
\label{fig:4_deltaF_depol_asymm}
\end{figure}

Thus far, we have considered identical gate-noise levels on Alice's and Bob's sides: $\lambda=\lambda_A=\lambda_B$. It is well known, and intuitively plausible, that nonlinear distillation protocols, such as the one studied here, can be sensitive to asymmetries between their input states. In particular, their performance may rely strongly on the two qubit pairs being prepared in the same input state. It is therefore important to investigate whether the protocol remains robust when the gate-error strengths on Alice's and Bob's sides differ. For a fixed preparation-noise parameter $\varepsilon$, we evaluate $\delta\mathcal F$ on a grid of depolarization strengths $\lambda_A,\lambda_B$ associated with Alice's and Bob's quantum computers, respectively. As shown in Fig.~\ref{fig:4_deltaF_depol_asymm}, the contour defined by $\delta\mathcal F=0$ is approximately aligned with the line $\lambda_A+\lambda_B=2\lambda^*(\varepsilon)$ for some $\lambda^*(\varepsilon)>0$. The overall structure of the diagram remains unchanged as $\varepsilon$ is varied: $\delta\mathcal F>0$ for $\lambda_A+\lambda_B<2\lambda^*(\varepsilon)$, and $\delta\mathcal F\leq0$ otherwise.

\begin{figure}
\centering
\includegraphics[width=.7\linewidth]{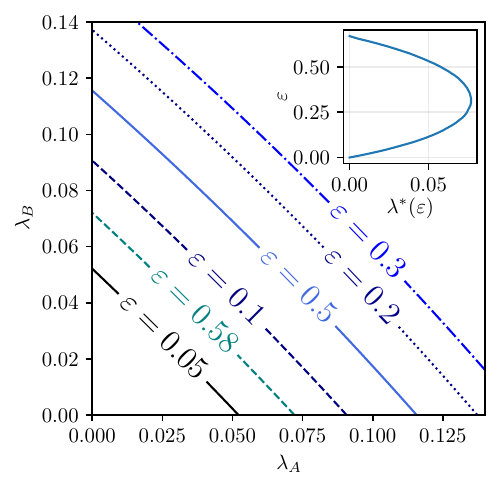}
\caption{$\delta\mathcal F=0$ curves on the ($\lambda_A, \lambda_B$) plane, corresponding to no enhancement in fidelity after four steps, in the presence of asymmetric depolarization errors, for different values of input white noise $\varepsilon$. The curves are aligned with $\lambda_A + \lambda_B = 2\lambda^*(\varepsilon)$. The inset in the upper-right corner shows $\lambda^*(\varepsilon)$ as a function of $\varepsilon$.}
\label{fig:5_c(epsilon)}
\end{figure}

We next examine the dependence of $\lambda^*(\varepsilon)$ on the preparation noise. To this end, we calculate $\delta\mathcal F$ for input states with different values of the white-noise parameter $\varepsilon$, while setting the depolarization strengths to $\lambda_A>0$, and $\lambda_B=0$. We then determine the value of $\lambda_A$ for which $\delta\mathcal F$ vanishes. This will give us $2\lambda^*(\varepsilon)$, given that the critical parameter curve is in fact aligned with $\lambda_A+\lambda_B=2\lambda^*$. The resulting dependence is shown in Fig.~\ref{fig:5_c(epsilon)}.

The curve describing $\lambda^*(\varepsilon)$ bears a clear resemblance to that of $\lambda_c(\varepsilon)$ shown in Fig.~\ref{fig:3_deltaF_depolarization}. This relation is natural given that the fidelity of the output state depends only on $\lambda_A + \lambda_B$ in first order. This property can be verified analytically for any noise channels which are of the form $\mathcal E^{\lambda_A}\otimes\mathcal E^{\lambda_B}$ (meaning that the same type of error occurs on Alice's and Bob's side, but with possibly different parameters). The detailed proof can be found in App.~\ref{app:B}. A very important consequence is that hereinafter we can restrict our analysis to symmetric errors, since any asymmetric channel with parameters $\lambda_A\neq\lambda_B$ is equivalent (up to first order) to a symmetric one with parameter $\lambda^* = (\lambda_A+\lambda_B)/2$.

Let us note that the proof in App.~\ref{app:B} assumes swap-invariant input states with respect to the qubits. As mentioned before, by restricting the input state to be of the form given in Eq.~\eqref{eq:rho0}, one loses generality. There might occur some interesting behavior when considering non-swap-invariant input states and highly asymmetric error channels, however these cases are not of interest of the current analysis.

\subsection{Amplitude damping}
\label{subsec:3B}
Another often considered gate noise model is amplitude damping, which accounts for transitions between qubit states. It is inherently described via a one-qubit quantum channel with the following Kraus operators:

\begin{equation}
    E_0(\gamma) = \begin{pmatrix}
        1 & 0 \\
        0 & \sqrt{1-\gamma}
    \end{pmatrix}, \qquad
    E_1(\gamma) = \begin{pmatrix}
        0 & \sqrt\gamma\\
        0 & 0
    \end{pmatrix},
\end{equation}
where $\gamma$ is the decay probability. $E_0$ corresponds to a ``no-decay'' event, in which the $\ket{1}$ state survives with probability $1-\gamma$, while $E_1$ describes a decay event, where with probability $\gamma$, the $\ket{1}$ state decays to $\ket{0}$.

When considering such channels, there are, in principle, four different $\gamma$ parameters, one for each qubit involved in the protocol. Since, in a physical architecture, the qubits are typically realized in the same manner, a simplified yet well-motivated assumption is to take all four parameters to be equal. Thus, our main interest will be in the interplay between the amplitude damping parameter and the strength of the preparation noise.

We note that the amplitude damping channel also exhibits the above-mentioned property that, to first order in the noise strength, a non-symmetric configuration ($\gamma_A \neq \gamma_B$) is equivalent to a symmetric one with an effective parameter
\[
\gamma_A'=\gamma_B'=\frac{\gamma_A+\gamma_B}{2},
\]
for $\gamma_A,\gamma_B\ll1$.

In our model, we may assume that on each party's side a two-qubit channel acts with the following Kraus operators:
\begin{equation}
    \mathcal K_{2i+j}(\gamma) = E_i(\gamma)\otimes E_j(\gamma), \quad i,j\in\{0,1\},
\end{equation}
so that the effect of the channel is given as
\begin{equation}
    \mathcal A^{(\gamma)}(\rho) = \sum_{k=0}^3 \mathcal K_k(\gamma) \,\rho\, \mathcal K_k^{\dagger}(\gamma).
\end{equation}

We will start with the calculation of the fidelity after each iteration of the protocol starting from inputs with white noise parameter $\varepsilon=0.2$, for different $\gamma$ values. Since the curves have a very similar tendency as those shown in the discussion of the depolarization error (Fig.~\ref{fig:2_depolarization_steps}), these results are presented in Fig.~\ref{fig:16_other_steps} in the Appendix. One important thing to observe is that for $\gamma$ values as low as $0.065$ the scheme already fails to distill states. We note however that $\gamma$ corresponds to a one-qubit decay, while $\lambda$ is the parameter of a two-qubit channel, thus comparing them in a one-to-one manner might be misleading.

\begin{figure}
\centering
\includegraphics[width=\linewidth]{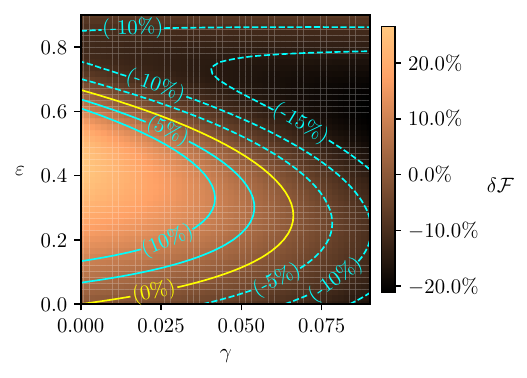}
\caption{Change in fidelity $\delta \mathcal F$ after four steps for input states with different white-noise parameters $\varepsilon$ and different amplitude damping parameters $\gamma$. The line corresponding to the critical parameter $\gamma_c(\varepsilon)$ is shown in yellow and satisfies $\delta\mathcal F=0$.}
\label{fig:7_deltaF_amplitude_damping}
\end{figure}

As a next step, we calculate the output fidelity after four iterations for different preparation noise and amplitude damping parameters. The results are displayed in Fig.~\ref{fig:7_deltaF_amplitude_damping}. Similarly to the case of depolarization, we can define a critical amplitude damping parameter $\gamma_c(\varepsilon)$. Starting from $\rho_0(\varepsilon)$, applying four distillation steps affected by an amplitude damping noise of  strength $\gamma_c(\varepsilon)$ does not lead to a fidelity enhancement (i.e., $\delta\mathcal F=0$), see the yellow curve in Fig.~\ref{fig:7_deltaF_amplitude_damping}. Below this, however, distillation can be achieved for relatively high values of $\varepsilon$.

\subsection{Bit flip and phase flip channels}
\label{subsec:3C}

\begin{figure*}
    \centering
    \includegraphics[width=0.49\linewidth]{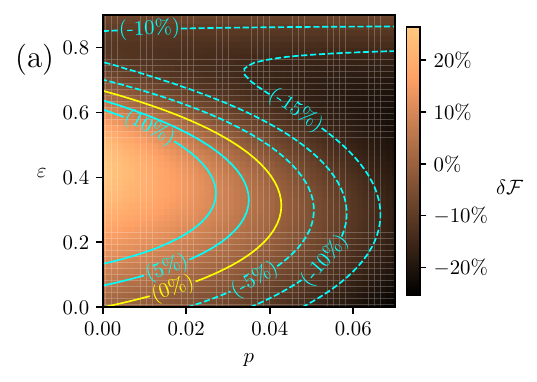}
    \hfill
    \includegraphics[width=0.49\linewidth]{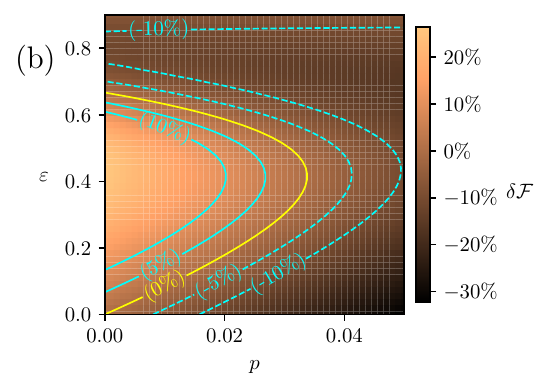}

    \caption{Change in fidelity $\delta \mathcal F$ after four iterations, in the presence of (a) \textit{bit flip} and (b) \textit{phase flip} errors. Each grid point corresponds to different input states $\rho_0(\varepsilon)$ and bit-flip and phase-flip parameters $p$ and $q$, respectively. The lines corresponding to the critical parameters $p_c(\varepsilon)$ and $q_c(\varepsilon)$ are shown in yellow.}
    \label{fig:11_phase_flip_deltaF}
\end{figure*}

In the next section we examine the effect of \textit{bit flip} and \textit{phase flip} channels on the feasibility and performance of the distillation scheme. Similarly to the case of amplitude damping, these errors are defined for single qubits, with Kraus operators
\begin{equation}
    E_0(p) = \sqrt{1-p}\,I, \quad E_1(p) = \sqrt{p}\,X
\end{equation}
for the case of \textit{bit flip}, and
\begin{equation}
    F_0(q) = \sqrt{1-q}\, I, \quad F_1(q) = \sqrt{q}\, Z
\end{equation}
for the case of \textit{phase flip.} From these operators the two-qubit channels can be built up via their operator elements:
\begin{equation}
    \mathcal{K}_{2i+j}(p) = E_i(p)\otimes E_j(p),
\end{equation}
and
\begin{equation}
    \mathcal L_{2i+j}(q) = F_i(q)\otimes F_j(q),
\end{equation}
respectively.

Using these the two-qubit channels are given by
\begin{equation}
    \mathcal X^{(p)}(\rho) = \sum_{k=0}^3 \mathcal K_{k}(p)\,\rho\,\mathcal K_k^{\dagger}(p)
\end{equation}
and
\begin{equation}
    \mathcal Z^{(q)}(\rho) = \sum_{l=0}^3 \mathcal L_l(q)\,\rho\,\mathcal L_l^{\dagger}(q).
\end{equation}

Note that we already built up the two-qubit channels in a symmetric way, with the parameters being the same for both qubits. This is the same simplification as the one we used in the case of amplitude damping as it can be shown that for moderate $p$ and $q$ parameters it is also sufficient to only consider  cases that are symmetric with respect to the exchange of the two parties; for details see App.~\ref{app:B}.

The changes in the fidelity as a function of the iterations for the case of bit-flip and phase-flip errors is presented in Fig.~\ref{fig:16_other_steps} of App.~\ref{app:D}, where one can see that the protocol proves far less robust against bit flip and phase flip errors than against depolarization and amplitude damping: parameters as low as 0.05 already cause the states to converge to the completely mixed state. Another feature to be noted is how fast the output fidelity drops when changing the channel parameter slightly. This shows that the protocol is more sensitive against phase flips than bit flips.

As in the previous cases, we examine the interplay between the initial noise and the error-channel parameters, as shown in Figs.~\ref{fig:11_phase_flip_deltaF} (a) and (b). The resulting diagrams exhibit a similar overall structure as in the case of the previously considered noise models. In each case, the critical parameter curve bounds a region of parameter space in which the protocol enhances the fidelity. In general, the greatest enhancement occurs for high preparation noise and low gate-error rates. However, the detailed features of the critical curves depend on the noise model. Most notably, the maximum tolerable error parameter is lower for the bit-flip and phase-flip models than for the other incoherent error models. Further properties of these diagrams will be examined in detail in the next section.

\subsection{Comparing incoherent noises}

\begin{figure}[h]
\centering
\includegraphics[width=.8\linewidth]{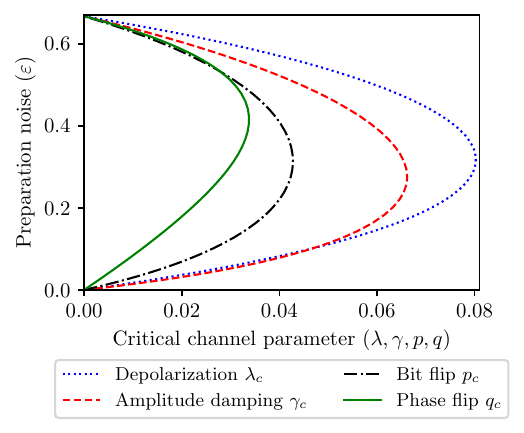}
\caption{Critical parameters satisfying $\delta \mathcal F=0$, for the different error channels as a function of the initial noise $\varepsilon$.}
\label{fig:12_incoherent_summary}
\end{figure}

We have seen so far that for incoherent noise channels a very similar overall tendency emerges. The critical parameters lie on a curve that intersects the $\varepsilon$ axis in zero and in some other $\varepsilon\neq0$ point. However the shape of these curves differ remarkably, so it remains challenging to characterize the robustness of the protocol with only one number. In order to summarize the results, we can plot the critical parameter curves on a single diagram, as shown in  Fig.~\ref{fig:12_incoherent_summary}. One can see that the protocol seems more robust against the depolarization and amplitude damping errors, while less robust in the case of bit flip and phase flip channels. One more quite important detail is worth noticing. In all cases, the critical parameter curves meet the $\varepsilon$ axis at exactly 2/3. This is a good sanity check as it is well known that for $\varepsilon\geq2/3$ the input state contains no entanglement, and with LOCC protocols it is impossible to create entanglement.

In order to characterize and to be able to compare the effect of these errors, in what follows, we introduce three quantities. The first one is the \textit{average fidelity enhancement} in the relevant region:

\newcommand{\D}{\mathrm{d}}

\begin{equation}
    \expval{\delta \mathcal F} = \dfrac{\int_{\mathcal P}\D\lambda\D\varepsilon\,  \delta \mathcal F(\lambda,\varepsilon)}{\int_{\mathcal P}\D\lambda\D\varepsilon},
\end{equation}
with
\begin{equation}
    \mathcal P = \{ (\lambda,\varepsilon) \,|\, \mathcal \delta F(\lambda,\varepsilon)\geq 0 \}.
\end{equation}

Another quantity that can offer key insights to the robustness against different noise types is the derivative of $\delta\mathcal F$ with respect to the noise parameter, at zero input noise, which we will call \textit{robustness} hereinafter:
\begin{equation}
    \mathcal R = \dfrac{\partial\;\delta\mathcal F(\lambda,\varepsilon)}{\partial \lambda}\bigg|_{\varepsilon,\lambda=0}.
\end{equation}

Finally, the maximum of the channel parameter along the critical parameter curve can also be insightful: above this value, there is no hope to ever distill entanglement under the influence of the given error (starting with white noise input states). We shall call this the \textit{bounding parameter}:
\begin{equation}
    \bar\lambda = \max_{\varepsilon\in[0,1]} \lambda_c(\varepsilon).
\end{equation}

The values of these quantities (calculated after four iterations) can be found for each type of incoherent error (as well as  coherent and readout errors, see later sections) in Table~\ref{tab:1}. Note that the robustness is closely connected to the derivative of the $\delta\mathcal F=0$ curve at the origin, which is transparent if one observes the ordering of the different models with respect their robustness. It can be seen that the average fidelity enhancement is approximately $10\%$ for each noise model, which is a promising result. Comparing the robustness measures, we find that depolarization and amplitude damping yield similar values of $\mathcal R$, both close to $-1$, whereas the bit-flip model, and especially the phase-flip model, exhibits a substantially lower value. This suggests that the protocol is more robust against depolarization and amplitude damping than against bit-flip and phase-flip errors. The bounding parameters also provide encouraging indications regarding the feasibility of the distillation scheme, as they all exceed $3\%$. Given that typical effective error parameters are well below $1\%$, these results further support the potential practical applicability of the protocol.

\begin{table*}[t]
        \centering
        \renewcommand{\arraystretch}{1.3}
        \begin{tabular}{|c||c|c|c|c||c|c||c|}
            \hline 
        Error channel
        & {\color{red}\makecell{Depolar-\\ization}}
        & {\color{red}\makecell{Amplitude\\Damping}}
        & {\color{red}\makecell{Bit\\Flip}}
        & {\color{red}\makecell{Phase\\Flip}}
        & {\color{blue}\makecell{Exchange\\Interaction}}
        & {\color{blue}\makecell{Phase\\Error}}
        & {\color{brown}\makecell{Balanced\\Readout}}
        \\ \hline\hline
         $\expval{\delta\mathcal F}$ & 10.09\,\% & 10.00\,\% & 10.09\,\% & 9.95\,\% & 12.70\,\% & 12.13\,\% & 8.65\,\% \\ \hline
         $\mathcal R$ & -1.25 & -1.00 & -2.00 & -6.00 & 0 & 0 & 0 \\ \hline 
         $\bar\lambda$ & 0.0803 & 0.0662 & 0.0428 & 0.0338 & 0.4115 & 0.5872 & 0.2181 \\ \hline\hline
         Parameter range & \multicolumn{4}{c||}{$\varepsilon\in[0,0.67], \;\lambda\in[0,0.085]$} & \multicolumn{2}{c||}{$\varepsilon\in[0,0.8], \;\lambda\in[0,\pi/4]$} & $\varepsilon\in[0,0.7], \;\lambda\in[0,0.5]$ \\ \hline
    \end{tabular}
    \caption{Average fidelity enhancement $\expval{\delta\mathcal F}$, robustness ($\mathcal R$), and bounding parameter ($\bar\lambda$) for different error models after four iterations. Red, blue, and brown colors mark incoherent, coherent, and readout errors, respectively. The values were calculated on a $101\times100$ grid spanning the parameter ranges specified in the last row. These ranges are reported to ensure reproducibility and to account for potential numerical errors. }
    \label{tab:1}
\end{table*}

\section{Introducing coherent noise}
\label{sec:4}

Apart from incoherent errors, coherent gate errors, such as over- or under-rotations, can also occur in quantum processors. When considering multiple qubits, coherent errors can be divided into two main classes: local and non-local ones. In our case, the overall gate error -- in terms of Alice and Bob -- should be local. On each side we will thus consider two different gate error types which are common in the literature: \textit{phase} error and \textit{exchange interaction} error. 
To include such errors in our protocol, we shall construct the corresponding noisy version of the two-qubit gate $U$ of Eq.~\eqref{eq:U}:
\begin{equation}
    U_{\rm n}(\theta) = E(\theta)\cdot U, \quad E(\theta) = e^{-i\theta H},
\end{equation}
where $H$ is some hermitian operator (the Hamiltonian corresponding to the noise) and $\theta\ll1$. The iteration step of Eq.~\eqref{eq:Omegaprime} then has to be modified accordingly:
\begin{equation}
    \Omega' = (U_{\rm n}(\theta_A)\otimes U_{\rm n}(\theta_B))\,\Omega\,(U_{\rm n}(\theta_A)\otimes U_{\rm n}(\theta_B))^{\dagger}.
\end{equation}

\subsection{Phase error}

The so-called phase error is generated by $Z\otimes Z$, so that
\begin{equation}
    E_{\rm phase}(\theta) = e^{-i\theta\,Z\otimes Z/2}.
\end{equation}

\begin{figure}
\centering
\includegraphics[width=\linewidth]{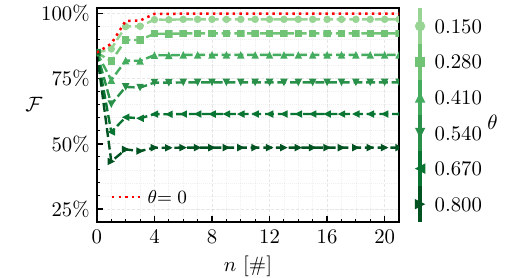}
\caption{Fidelity of the output state with respect to $\ket{\Phi_+}$ after successive iterations of the protocol. The input state corresponds to $\varepsilon=0.2$, and each curve represents a different value of the coherent phase error parameter ($\theta=\theta_A=\theta_B$).}
\label{fig:9_coherent_phase_symm_steps}
\end{figure}

In accordance with Appendix \ref{app:B}, we only consider the symmetric case with $\theta_A = \theta_B$. The fidelity of the output after a few iterations, starting from the previously considered input with 20\% white noise, is displayed in Fig.~\ref{fig:9_coherent_phase_symm_steps}.
An interesting behavior emerges: Even at relatively high values of $\theta$, the protocol seems not to break down; the output fidelity stays well over 0.25, unlike in the case of incoherent errors. On the other hand, not every $\theta$ leads to an enhancement of the fidelity, hence it is useful to calculate $\delta\mathcal F$ again.

\begin{figure*}         
\centering
\includegraphics[width=.48\linewidth]{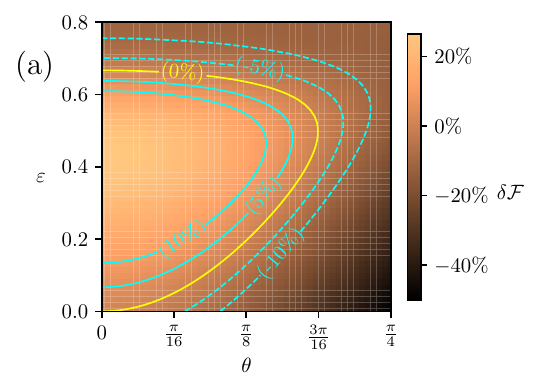}\hfill \includegraphics[width=0.48\linewidth]{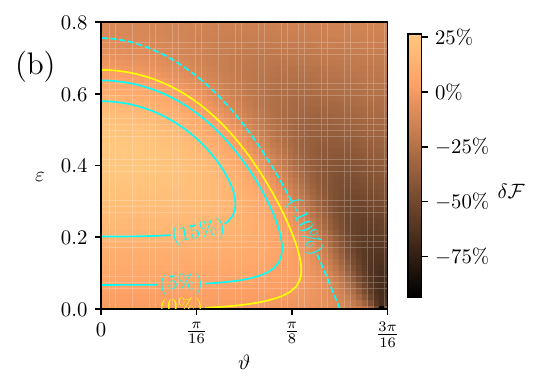} \caption{Change in fidelity $\delta \mathcal F$ after four iterations, in the presence of (a) \textit{phase error} and (b) \textit{exchange interaction} error. Each grid point corresponds to different input states $\rho_0(\varepsilon)$ and  phase or exchange interaction parameters, respectively. The lines corresponding to the critical parameters $\theta_c(\varepsilon)$ and  $\vartheta_c(\varepsilon)$,  are shown in yellow.}
\label{fig:17_coherent_exchange_symm_deltaF}
\end{figure*}

Figure~\ref{fig:17_coherent_exchange_symm_deltaF}(a) shows $\delta\mathcal{F}$ as a function of the initial noise and the coherent phase-error parameter $\theta$, following the same approach as for incoherent errors. The overall trend is similar: the critical error curve encloses a region in parameter space where the protocol enhances the fidelity, although its shape differs from those obtained for incoherent noise. At $\varepsilon=\theta=0$, the curve appears to have a vanishing derivative, which can be shown straightforwardly. This behavior arises because $\delta\mathcal{F}$ has no first-order correction in $\theta$, resulting in a robustness measure of zero. This value is reported in Table~\ref{tab:1}, alongside the results for the other noise models. The bounding parameter is almost $0.6$ rad, which is remarkably large compared with physically relevant error magnitudes. Finally, the average fidelity enhancement exceeds $12\%$, providing further evidence of the protocol's potential.

\subsection{Exchange interaction error}

Another frequently considered coherent gate error is the so-called exchange interaction error, relevant, e.g., for spin qubits, where the exchange interaction has also been proposed as the entangling operation of purification protocols~\cite{AuerBurkard2014}:
\begin{equation}
    E_{\rm exchange}(\vartheta) = e^{-i\vartheta\,(X\otimes X + Y\otimes Y)/2}.
\end{equation}

\begin{figure}
\centering
\includegraphics[width=\linewidth]{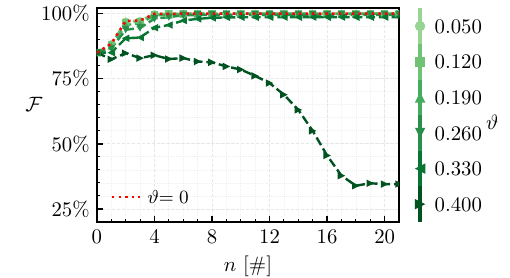}
\caption{Fidelity of the output state with respect to $\ket{\Phi_+}$ after successive protocol iterations. The input state
corresponds to $\varepsilon=0.2$, and each curve represents a different value of the coherent exchange interaction error parameter $\vartheta=\vartheta_A=\vartheta_B$.}
\label{fig:11_coherent_exchange_symm_steps}
\end{figure}

Similarly to the case of the phase error, we start by calculating the fidelity after every iteration for different $\vartheta=\vartheta_A = \vartheta_B$ parameters. This is displayed in Fig.~\ref{fig:11_coherent_exchange_symm_steps}, where a peculiar behavior emerges: for small $\vartheta$, after about 6-8 iterations, every case results in an output with a relatively high fidelity, which does not depend so strongly on the value of $\vartheta$ as in the other cases above. For higher noise levels, however, the protocol slowly fails, but the output is not the maximally mixed state since the fidelity stays above 0.25. 

Moving on to the interplay between the preparation noise and the gate error, let us examine the change in the fidelity on the $\varepsilon-\vartheta$ plane in Fig.~\ref{fig:17_coherent_exchange_symm_deltaF} (b). The above-mentioned behavior (i.e., the insensitivity to the value of $\vartheta$) puts a mark on this diagram. Near the critical parameter curve, lies a region with a quite sharp slope in which the fidelity drops significantly.

In order to compare the effect of these coherent errors, we calculated the \textit{average fidelity enhancement}, the \textit{robustness} and the \textit{bounding parameter} (see Table~\ref{tab:1}) similarly to the case of incoherent errors. The robustness is exactly zero again, as in the case of the phase error. The average fidelity enhancement is almost $13\%$, and the bounding parameter is also quite high, $0.41$ rad. These values show that the protocol remains feasible in the presence of moderate exchange interaction errors.

\section{Readout errors}
\label{sec:5}

Since the distillation scheme is based on intermediate measurements and post-selection, it is expected that errors in the measurements can lead to the breakdown of the protocol. In this section we will consider incoherent readout errors. These are essentially described by two parameters: the probabilities of false-readouts, $P(\tilde1|0)$ and $P(\tilde0|1)$. The former gives the probability with which a qubit in state $\ket0$ gives 1 when measured, and vice versa. In a simple model these probabilities may be assumed to be equal, $P(\tilde1|0)=P(\tilde0|1)$. Hereinafter this case will be referred to as "balanced error", while the case $P(\tilde1|0)\neq P(\tilde0|1)$ will be called "biased error". These terms are not to be confused with the properties "symmetric" and "asymmetric", which correspond to the exchange of Alice and Bob.

\subsection{Balanced error}
Balanced readout error must be included in the model similarly to incoherent gate errors. First of all, we shall construct the corresponding quantum channel:
\begin{equation}
    \mathcal{R}^{(p)}(\rho) = (1-p)\,Q_0\rho Q_0 + p\, Q_1\rho Q_1,
\end{equation}
with $Q_0=\id\otimes\ket0\bra0$ and $Q_1=\id\otimes\ket1\bra1$. Here we used the simplified notation $p=P(\tilde1|0)$. Applying this channel on an arbitrary input state results in the state of the system given that the faulty measurement reported the outcome 0. Using this, one iteration step of the protocol can be written as
\begin{equation}
    \rho' = \Tr_{2,4}\{ [\mathcal{R}^{(p_A)}\otimes \mathcal{R}^{(p_B)}](\Omega') \},
\end{equation}
followed by proper normalization. 

\begin{figure}
\centering
\includegraphics[width=\linewidth]{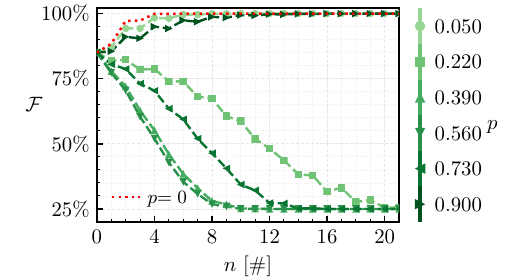}
\caption{Fidelity of the output state with respect to $\ket{\Phi_+}$ after successive protocol iterations. The input state corresponds to $\varepsilon=0.2$, and each curve represents a different value of the balanced symmetric readout error parameter $p$.}
\label{fig:12_readout_balanced_symm_steps}
\end{figure}

\begin{figure}
\centering
\includegraphics[width=\linewidth]{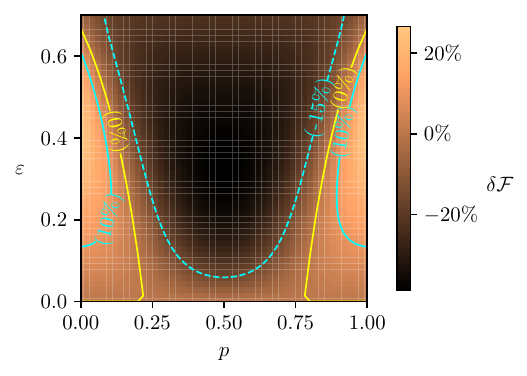}
\caption{Change in fidelity $\delta \mathcal F$ after four steps for input states with different white-noise parameters $\varepsilon$ and different balanced symmetric readout error parameters $p$. The line corresponding to the critical parameter $p_c(\varepsilon)$ is shown in yellow.}
\label{fig:21_readout_balanced_symm_deltaF}
\end{figure}

Starting with the symmetric case ($p_A=p_B$) we calculate the output fidelity after each iteration of the protocol, for a wide range of $p$ values (see Fig.~\ref{fig:12_readout_balanced_symm_steps}).
An interesting phenomenon emerges, namely the protocol seems to work flawlessly at sufficiently low and sufficiently high error parameters, while it tends to fail at parameters in between. To investigate this behavior further, we calculate the change in fidelity after four iterations for a wide range of $p$ and $\varepsilon$ values to see the interplay between readout and preparation errors, as shown in Fig.~\ref{fig:21_readout_balanced_symm_deltaF}. From this, one can extract the critical parameter curve $p_c(\varepsilon)$. The results clearly show that the performance of the protocol is in fact symmetric in $p$: $\delta \mathcal F$ is the same for $p$ and for $1-p$. This suggests that for certain input states the post-selection on the '$11$' measurement outcome may lead to the same output as the post-selection on the '$00$' case. This would mean that the success probability of the distillation can be straightforwardly enhanced by accepting both strings. This behavior will be further investigated in section~\ref{sec:7}. One can find the corresponding \textit{average fidelity enhancement}, the \textit{robustness} and the \textit{bounding parameter} in Table~\ref{tab:1}.\\

\begin{figure}
\centering
\includegraphics[width=\linewidth]{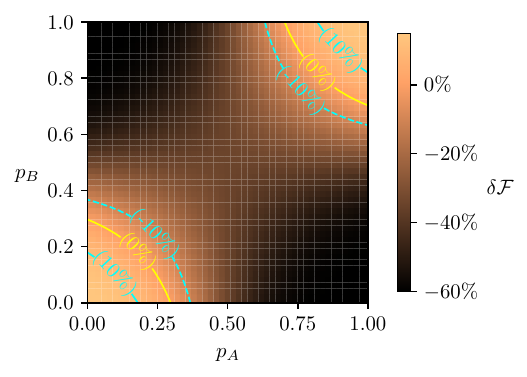}
\caption{Change in fidelity $\delta \mathcal F$ after four steps for a fixed input state with white-noise parameter $\varepsilon=0.2$ and different balanced faulty reading probabilities on Alice's and Bob's sides, respectively. The line corresponding to $\delta \mathcal F=0$ is highlighted in yellow.}
\label{fig:22_readout_balanced_asymm_deltaF}
\end{figure}

Turning our attention to the asymmetric case, we calculate the change in fidelity after four steps of the protocol, given that the faulty readout probabilities ($p_A$ and $p_B$) are not necessarily equal on Alice's and Bob's side. The results are shown in Fig.~\ref{fig:22_readout_balanced_asymm_deltaF}. We note that for $p_A,p_B\ll1$, the usual property holds for the output fidelity, namely any asymmetric setup is equivalent to a symmetric one. However, since in this case a much wider region of parameters is enclosed by the critical curve, it is interesting to see how the full parameter-space looks like.
The protocol seems to fail in those regions where $p_A$ and $p_B$ differ the most. One can also see that the diagram is symmetric under Alice's and Bob's exchange as well as under the simultaneous exchanges $p_A\leftrightarrow 1-p_A$ and  $p_B\leftrightarrow 1-p_B$.

\subsection{Biased error}
Now we turn to the problem of biased errors, i.e. readout errors where $P(\tilde1|0)\neq P(\tilde0|1)$. For the sake of simplicity, let us use the notations $p=P(\tilde1|0)$, and $q=P(\tilde0|1)$. The corresponding map can be constructed similarly to the previous case:
\begin{equation}
    \mathcal{R}^{(p,q)}(\rho) = (1-p)\,Q_0\rho Q_0 + q\, Q_1\rho Q_1.
\end{equation}

For completeness, we also present the operation corresponding to the reported outcome 1:
 \begin{equation}
    \mathcal{P}^{(p,q)}(\rho) = p\,Q_0\rho Q_0 + (1-q)\, Q_1\rho Q_1.
\end{equation}
The pair $\{\mathcal{R}^{(p,q)},\mathcal{P}^{(p,q)}\}$ describes a binary readout channel with asymmetric assignment errors, parameterized by the probabilities \(p\) and \(q\).

\begin{figure}
    \centering
    \includegraphics[width=\linewidth]{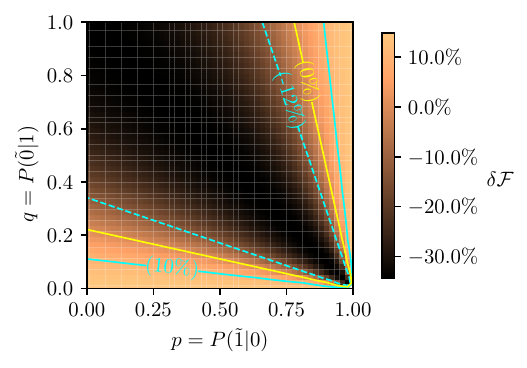}
    \caption{The change in fidelity $\delta\mathcal F$ for different readout noise probabilities $p$ and $q$, considering a symmetric biased model, and a fixed preparation noise corresponding to $\varepsilon=0.2$. The yellow line represents the error values for which $\delta\mathcal F=0$.}
    \label{fig:15_readout_biased_symm}
\end{figure}

Using the above, one can get the unnormalized state after one iteration as
\begin{equation}
    \rho' = \Tr_{2,4}\{ [\mathcal{R}^{(p_A, q_A)}\otimes \mathcal{R}^{(p_B,q_B)}](\Omega') \}.
\end{equation}

One can see that there are four parameters describing such errors: Two on Alice's and two on Bob's side. We shall now restrict our attention to the symmetric case, where $p_A=p_B$ and $q_A=q_B$. We can calculate the change in fidelity $\delta\mathcal F$ for different $p,q$ values sampled from a grid, while keeping the preparation noise parameter $\varepsilon$ constant, as shown in Fig.~\ref{fig:15_readout_biased_symm}. It can be seen that the plot is symmetric to the simultaneous $p\leftrightarrow1-p$ and $q\leftrightarrow 1-q$ substitutions. This aligns well with the observation made in the balanced symmetric case: For well-behaved input states, the '$00$' and the '$11$' post-selection leads to similar dynamics. A more interesting feature of the plot is that the protocol seems to be more robust against the $p$-type errors than the $q$-type ones. This is not surprising, considering the following arguments. Setting $q$ to zero while increasing $p$ means that the $\ket1$ states of qubits $2$ and $4$ are read as $1$, but the $\ket0$ states have a higher and higher probability to be heralded as $1$ as well. In such a false readout, the remaining qubits (1,3) are simply thrown away, not affecting the output at all. On the other hand, increasing $q$ corresponds to a faulty readout of $\ket1$ states as '0' with a higher probability. This leads to accepting "incorrect measurements", such as '$01$', '$10$', or '$11$'; which then lowers the output fidelity.

\section{Examples}
\label{sec:6}

So far, we have considered each type of gate error separately. In a more realistic setting, however, multiple sources of error may occur simultaneously. In this section, we present several examples illustrating the performance of the protocol under such conditions, taking into account preparation errors, incoherent and coherent gate noise, and readout errors. The selected parameter values lie within the critical curves identified in the preceding analysis, ensuring that each individual error parameter remains in a regime where distillation is expected to be beneficial.

Table~\ref{tab:3} summarizes the results. The examples are organized into four blocks, each corresponding to a different incoherent noise channel: depolarization, amplitude damping, bit flip, and phase flip. Within each block, the two rows represent different combinations of preparation noise, coherent gate errors, and readout errors. In all cases, preparation noise is present alongside one incoherent noise channel, one coherent error model, and readout errors. This allows us to examine whether the protocol remains effective when several distinct imperfections act simultaneously.

The error parameters were selected manually to illustrate representative scenarios rather than to reproduce a specific experimental implementation. In particular, the preparation-error parameters are of the order of $10^{-1}$, whereas the incoherent gate-noise parameters are smaller, typically of the order of $10^{-3}$--$10^{-2}$. These values are not directly calibrated against experimental data. Instead, we deliberately chose relatively large error parameters to test whether distillation can remain feasible even under appreciable imperfections. The results should therefore be interpreted as illustrative examples of the protocol's robustness, rather than as quantitative predictions for a particular physical platform.

The results suggest that the protocol remains feasible even in the presence of multiple error sources. All combinations except one yield a substantial fidelity enhancement of between $11$ and $20$ percentage points, which is promising. The smallest enhancement ($3.7$ percentage points) occurs for the input with the lowest preparation noise ($\varepsilon=0.1$), where the input fidelity is already high; the output fidelity ($96.2\%$) is comparable to that of the other cases.

\begin{table*}[t]
    \centering
    {\scriptsize
    \begin{tabular}{|c|c|c|c|c|c|c|c||c|c|c|}
        \hline
         {\color{orange} Prep $(\varepsilon)$} & {\color{red}Depol. $(\lambda)$} & {\color{red}Amp. Damp. $(\gamma)$} & {\color{red}Bit Flip $(p)$} & {\color{red}Phase Flip $(q)$} & {\color{blue}Ex. Int. $(\vartheta)$} & {\color{blue}Phase $(\theta)$} & {\color{brown}Readout $(p)$} & $\mathcal F_{\rm in}\,\%$& $\mathcal F_{\rm out}\,\%$& $\delta\mathcal F\,\%$\\\hline\hline
         
         0.35 & $0.002$ & - & - & - & $\pi/16$ & - & 0.005 & 73.75& 93.04& \textbf{19.29}\\\hline
         
         0.25 & 0.01 & - & - & - & - & $\pi/16$ & 0.008 & 81.25 & 93.89& \textbf{12.64}\\\hline\hline
         
         0.35 & - & 0.02 & - & - & $\pi/20$ & - & 0.004 & 73.75 & 87.97 & \textbf{14.22}\\\hline
         
         0.28 & - & $0.004$ & - & - & - & $\pi/20$ & 0.01 & 79 & 95.76 & \textbf{16.76}\\\hline\hline
         
         0.3 & - & - & $0.001$ & - & $\pi/15$ & - & 0.01 & 77.5 & 94.66 & \textbf{17.16} \\\hline
         
         0.2 & - & - & $0.003$ & - & - & $\pi/25$ & 0.005 & 85 & 97.45 & \textbf{12.45} \\\hline\hline
         
         0.1 & - & - & - & 0.005 & $\pi/20$ & - & 0.01 & 92.5 & 96.23 & \textbf{3.73}\\\hline
         
         0.2 & - & - & - & 0.003 & - & $\pi/25$ & 0.005 & 85 & 96.28 & \textbf{11.28}\\\hline
    \end{tabular}}
    \caption{Examples involving several types of error models simultaneously. For each case, the preparation noise (marked in orange) was taken nonzero, and one of the incoherent noises (marked in red), and one of the coherent errors (marked in blue) were considered alongside with a given readout error (marked in brown). The input and the output fidelity $\mathcal{F}_{in}$ and $\mathcal{F}_{out}$, as well as their difference $\delta\mathcal{F}$, are presented in the last columns. The different error parameters were chosen to serve as rough order-of-magnitude estimates.}
    \label{tab:3}
\end{table*}

\section{Enhancing success probabilities}
\label{sec:7}

The analysis of balanced symmetric readout errors suggests a possible improvement of the protocol, as discussed in Section~\ref{sec:5}. Specifically, for the input state considered here and in the absence of gate errors, it appears that post-selection based on accepting the '$00$' measurement outcome and post-selection based on accepting the '$11$' outcome lead to the same dynamics. 
One might argue that the figures presented above show only the output fidelity, which itself exhibits the aforementioned symmetry, and therefore do not establish that the underlying states possess the same symmetry. However, it is important to note that the fidelity was evaluated only after four iterations of the protocol. If the states produced by the two post-selection schemes were not similar, their respective dynamics could progressively diverge over successive iterations.

Our next objective is therefore to formulate this apparent symmetry mathematically and to identify the conditions on the input state and gate noise under which the two post-selection schemes give rise to identical dynamics.

\subsection{Conditions for the input state}
Starting from a general input state, it is straightforward to calculate the unnormalized output state, considering the post-selection for the '$ii\,$' measurement output, see the detailed calculations in App.~\ref{app:C}. Denoting this state with $\rho_i'$, we can postulate two conditions which guarantee the same dynamics in the two cases. First, $\rho_0' = \rho_1'$ must hold, which gives some equations for the input state. Solving these we get $[\rho,X\otimes X]=0$. This means that if the input state commutes with $X\otimes X$, then both the '$00$' and the '$11$' post-selection leads to the same output state. This is not enough however, as we must ensure that the resulting state also meets the conditions, thus we need $[\rho_0',X\otimes X]=[\rho_1',X\otimes X]=0$. This gives another set of equations for the input state, which turns out to be equivalent to the condition $[\rho,Z\otimes Z]=0$. Thus we get that the input states satisfying
\begin{equation}
\label{eq:rho_belldiag_condition}
    [\rho,X\otimes X] = [\rho,Z\otimes Z]=0
\end{equation}
give rise to the same dynamics for both the '$00$' and the '$11$' post-selections. These conditions are met if and only if the input state is \textit{Bell-diagonal}. This is in line with the recurrence protocols of Refs.~\cite{Bennett_distill_1996,Deutsch_distill_1996}, which operate on Werner and Bell-diagonal states, respectively, and accept both coinciding measurement outcomes. 

\subsection{Conditions for the noise channel}
In App.~\ref{app:C_2} we derive the equations for the noise channel that need to be met in order for the '$00$' and '$11$' post-selections to give the same dynamics. 
To this end, we start from a Bell-diagonal state (since we do not expect the protocol to admit a wider class of input states in the presence of errors), and first impose that the output state should be Bell-diagonal as well. 
Next, assuming that these hold, we impose that the output state is symmetric under the '$00$'-'$11$' post-selection. After solving the equations corresponding to these conditions 
we find that the depolarizing, bit-flip, and phase-flip channels considered in this work preserve both Bell diagonality and the $00$--$11$ symmetry. By contrast, amplitude damping and the coherent error models considered in Sec.~IV do not in general satisfy these conditions. The results are summarized in table~\ref{tab:2}.

For channels satisfying these requirements, both '$00$' and '$11$' may therefore be retained without changing the conditional distillation dynamics, thereby increasing the success probability of each iteration.
 This property can be really useful, considering the probabilistic nature of the protocol. Each iteration step has a success probability, which can be enhanced when accepting both the '$00$' and the '$11$' outcomes. This becomes even more relevant, if one wants to calculate the rate at which they can distill Bell-states. For one iteration, two pairs are needed. For two iterations, four input states are needed, and in total 3 pairs of measurements. For $n$ iterations, the number of measurements scales exponentially, and thus the deeper the circuit, the more we gain in distillation rate, when accepting both outcomes. Since for Bell-diagonal inputs and such channels the unnormalized output states coincide, the two outcomes also occur with equal probability. Accepting both therefore doubles the success probability of every iteration, i.e., increases the overall success probability by a factor of $2^n$ after $n$ iterations. For example, for $\varepsilon=0.2$ and four ideal iterations, the expected number of input pairs consumed per output pair decreases from about $422$ to about $26$.

\newcommand{\cmark}{\ding{51}}
\newcommand{\xmark}{\ding{55}}

\begin{table}[h]
    \centering
    \begin{tabular}{|c|c|c|}
        \hline
        Channel & \makecell{Preserves\\Bell-diagonality} & \makecell{00-11\\ symmetry} \\\hline\hline
        {\color{red} Depolarization} & \cmark & \cmark\\\hline
        {\color{red} Bit flip} & \cmark & \cmark \\\hline
        {\color{red} Phase flip} & \cmark & \cmark \\\hline
        {\color{red} \makecell{Amplitude\\damping}} & \xmark & \xmark \\\hline
        {\color{red} \makecell{Diagonal\\PTM}} & \cmark & \cmark \\\hline\hline
        {\color{blue} \makecell{Exchange\\interaction}} & \xmark & \xmark \\\hline
        {\color{blue} Phase error} & \xmark & \xmark \\\hline
    \end{tabular}
    \caption{Properties of incoherent and coherent error channels regarding the '$00$'-'$11$' symmetry conditions.}
    \label{tab:2}
\end{table}

\section{Conclusion}
\label{sec:8}

In this work, we analyzed a probabilistic LOCC protocol for Bell-pair distillation, focusing on its feasibility under imperfect operations described by different noise models. We showed that, for moderate gate errors, the protocol can enhance the fidelity of the distributed qubit pairs after four consecutive iterations. We identified the parameter ranges within which the protocol remains feasible and found that, for the depolarizing noise model, currently available devices fall within this range, highlighting the potential of the scheme for practical entanglement distillation. For incoherent gate noise, the tolerable error levels are of the order of a few percent, comparable to the thresholds reported for recurrence-type protocols with noisy gates~\cite{Dur_Briegel_1999,Rohde2007}. A quantitative comparison, however, would require evaluating the protocols under the same error model and figure of merit, since the parameters of different noise channels are not directly equivalent. 

Finally, we established criteria under which complementary postselection outcomes can be exploited to increase the success probability, further improving the protocol's practical feasibility. For Bell-diagonal inputs and Pauli-type gate noise, accepting both outcomes, as in the recurrence protocols~\cite{Bennett_distill_1996,Deutsch_distill_1996}, doubles the success probability of every iteration. Natural next steps are platform-specific noise models, e.g., for color centers~\cite{Alessio_2026} and superconducting circuits, where entanglement distillation has already been demonstrated~\cite{Kalb2017,Yan2022}, and the combination of the protocol with the purification of noisy measurements~\cite{KimYunBae2025}.

\section*{Acknowledgments}
This work was supported by the European Union’s Horizon Europe research and innovation program under grant agreement No 101135699 via the project SPINUS, and the Horizon Europe programme HORIZON-CL4-2022-QUANTUM-01-SGA via the project 101113946 OpenSuperQPlus100. 
We acknowledge support from the National Research, Development and Innovation Office of Hungary, project No. TKP-2021-NVA-04. TK is grateful for the support from the ‘Frontline’ Research Excellence Programme of the NKFIH (Grant No. KKP133827). AK also acknowledges support from the ‘EKÖP’ Research Programme of ELTE (Grant No. EKÖP-26-2-I-ELTE-20260576).

\bibliographystyle{apsrev4-2}
\bibliography{journals}

@article{Bennett1993,
  title = {Teleporting an unknown quantum state via dual classical and Einstein-Podolsky-Rosen channels},
  author = {Bennett, Charles H. and Brassard, Gilles and Cr\'epeau, Claude and Jozsa, Richard and Peres, Asher and Wootters, William K.},
  journal = {Phys. Rev. Lett.},
  volume = {70},
  issue = {13},
  pages = {1895--1899},
  numpages = {0},
  year = {1993},
  month = {Mar},
  publisher = {American Physical Society},
  doi = {10.1103/PhysRevLett.70.1895},
  url = {https://link.aps.org/doi/10.1103/PhysRevLett.70.1895}
}

@article{Bennett_distill_1996,
  title = {Purification of Noisy Entanglement and Faithful Teleportation via Noisy Channels},
  author = {Bennett, Charles H. and Brassard, Gilles and Popescu, Sandu and Schumacher, Benjamin and Smolin, John A. and Wootters, William K.},
  journal = {Phys. Rev. Lett.},
  volume = {76},
  issue = {5},
  pages = {722--725},
  numpages = {0},
  year = {1996},
  month = {Jan},
  publisher = {American Physical Society},
  doi = {10.1103/PhysRevLett.76.722},
  url = {https://link.aps.org/doi/10.1103/PhysRevLett.76.722}
}

@article{Bennett_distill_2_1996,
  title = {Mixed-state entanglement and quantum error correction},
  author = {Bennett, Charles H. and DiVincenzo, David P. and Smolin, John A. and Wootters, William K.},
  journal = {Phys. Rev. A},
  volume = {54},
  issue = {5},
  pages = {3824--3851},
  numpages = {0},
  year = {1996},
  month = {Nov},
  publisher = {American Physical Society},
  doi = {10.1103/PhysRevA.54.3824},
  url = {https://link.aps.org/doi/10.1103/PhysRevA.54.3824}
}

@article{Deutsch_distill_1996,
  title = {Quantum Privacy Amplification and the Security of Quantum Cryptography over Noisy Channels},
  author = {Deutsch, David and Ekert, Artur and Jozsa, Richard and Macchiavello, Chiara and Popescu, Sandu and Sanpera, Anna},
  journal = {Phys. Rev. Lett.},
  volume = {77},
  issue = {13},
  pages = {2818--2821},
  numpages = {0},
  year = {1996},
  month = {Sep},
  publisher = {American Physical Society},
  doi = {10.1103/PhysRevLett.77.2818},
  url = {https://link.aps.org/doi/10.1103/PhysRevLett.77.2818}
}

@article{Macchiavello_1998,
   title={On the analytical convergence of the QPA procedure},
   volume={246},
   ISSN={0375-9601},
   url={http://dx.doi.org/10.1016/S0375-9601(98)00550-7},
   DOI={10.1016/s0375-9601(98)00550-7},
   number={5},
   journal={Physics Letters A},
   publisher={Elsevier BV},
   author={Macchiavello, Chiara},
   year={1998},
   month=sep, pages={385–388} }

@article{Dur_Briegel_1998,
  title = {Quantum Repeaters: The Role of Imperfect Local Operations in Quantum Communication},
  author = {Briegel, H.-J. and D{\"u}r, W. and Cirac, J. I. and Zoller, P.},
  journal = {Phys. Rev. Lett.},
  volume = {81},
  issue = {26},
  pages = {5932--5935},
  numpages = {0},
  year = {1998},
  month = {Dec},
  publisher = {American Physical Society},
  doi = {10.1103/PhysRevLett.81.5932},
  url = {https://link.aps.org/doi/10.1103/PhysRevLett.81.5932}
}

@article{Dur_Briegel_1999,
   title={Quantum repeaters based on entanglement purification},
   volume={59},
   ISSN={1094-1622},
   url={http://dx.doi.org/10.1103/PhysRevA.59.169},
   DOI={10.1103/physreva.59.169},
   number={1},
   journal={Physical Review A},
   publisher={American Physical Society (APS)},
   author={Dür, W. and Briegel, H.-J. and Cirac, J. I. and Zoller, P.},
   year={1999},
   month=Jan, pages={169–181} }

@article{Hsieh_2004,
   title={A simplification of entanglement purification},
   volume={328},
   ISSN={0375-9601},
   url={http://dx.doi.org/10.1016/j.physleta.2004.06.005},
   DOI={10.1016/j.physleta.2004.06.005},
   number={2-3},
   journal={Physics Letters A},
   publisher={Elsevier BV},
   author={Hsieh, Jin-Yuan and Li, Che-Ming and Chuu, Der-San},
   year={2004},
   month=jul, pages={94–101} }

@article{Torres_2016,
  title = {Conditions for entanglement purification with general two-qubit states},
  author = {Torres, Juan Mauricio and Bern{\'a}d, J{\'o}zsef Zsolt},
  journal = {Phys. Rev. A},
  volume = {94},
  issue = {5},
  pages = {052329},
  numpages = {7},
  year = {2016},
  month = {Nov},
  publisher = {American Physical Society},
  doi = {10.1103/PhysRevA.94.052329},
  url = {https://link.aps.org/doi/10.1103/PhysRevA.94.052329}
}

@article{Rozpedek_2018,
   title={Optimizing practical entanglement distillation},
   volume={97},
   ISSN={2469-9934},
   url={http://dx.doi.org/10.1103/PhysRevA.97.062333},
   DOI={10.1103/physreva.97.062333},
   number={6},
   journal={Physical Review A},
   publisher={American Physical Society (APS)},
   author={Rozpedek, Filip and Schiet, Thomas and Thinh, Le Phuc and Elkouss, David and Doherty, Andrew C. and Wehner, Stephanie},
   pages={062333},
   year={2018},
   month=jun }

@article{Bernad_2022,
  title = {Optimal two-qubit gates in recurrence protocols of entanglement purification},
  author = {Preti, Francesco and Calarco, Tommaso and Torres, Juan Mauricio and Bern{\'a}d, J{\'o}zsef Zsolt},
  journal = {Phys. Rev. A},
  volume = {106},
  issue = {2},
  pages = {022422},
  numpages = {11},
  year = {2022},
  month = {Aug},
  publisher = {American Physical Society},
  doi = {10.1103/PhysRevA.106.022422},
  url = {https://link.aps.org/doi/10.1103/PhysRevA.106.022422}
}

@misc{Marty2026,
      title={Practical Entanglement Distillation Protocols with Quadratic Error Suppression}, 
      author={B{\"a}umer Marty, Elisa},
      year={2026},
      eprint={2605.26757},
      archivePrefix={arXiv},
      primaryClass={quant-ph},
      url={https://arxiv.org/abs/2605.26757}, 
}

@article{Kalman2025,
  title = {Universal, Unambiguous Concentration and Distillation of Bell pairs},
  author = {K\'alm\'an, Orsolya and G\'abris, Aur\'el and Jex, Igor and Kiss, Tam\'as},
  journal = {Phys. Rev. Lett.},
  volume = {135},
  issue = {26},
  pages = {260202},
  numpages = {6},
  year = {2025},
  month = {Dec},
  publisher = {American Physical Society},
  doi = {10.1103/rcx1-w6j7},
  url = {https://link.aps.org/doi/10.1103/rcx1-w6j7}
}

@article{Orthey_2026,
  title = {Optimality of universal conclusive entanglement concentration protocols},
  author = {Orthey, Alexandre C. and Philip, Aby and Kondra, Tulja Varun and Streltsov, Alexander},
  journal = {Phys. Rev. A},
  volume = {113},
  issue = {6},
  pages = {062404},
  numpages = {8},
  year = {2026},
  month = {Jun},
  publisher = {American Physical Society},
  doi = {10.1103/bz3j-9njg},
  url = {https://link.aps.org/doi/10.1103/bz3j-9njg}
}

@book{NielsenChuang2010, place={Cambridge}, title={Quantum Computation and Quantum Information: 10th Anniversary Edition}, publisher={Cambridge University Press}, author={Nielsen, Michael A. and Chuang, Isaac L.}, year={2010}}

@article{Pompili2022,
  author = {Pompili, M. and Delle Donne, C. and te Raa, I. and van der Vecht, B. and Skrzypczyk, M. and Ferreira, G. and de Kluijver, L. and Stolk, A. J. and Hermans, S. L. N. and Pawelczak, P. and Kozlowski, W. and Hanson, R. and Wehner, S.},
  title = {Experimental demonstration of entanglement delivery using a quantum network stack},
  journal = {npj Quantum Information},
  volume = {8},
  pages = {121},
  year = {2022},
  doi = {10.1038/s41534-022-00631-2}
}

@article{Bartling2025,
  author = {Bartling, H. P. and Yun, J. and Schymik, K. N. and van Riggelen, M. and Enthoven, L. A. and van Ommen, H. B. and Babaie, M. and Sebastiano, F. and Markham, M. and Twitchen, D. J. and Taminiau, T. H.},
  title = {Universal high-fidelity quantum gates for spin qubits in diamond},
  journal = {Physical Review Applied},
  volume = {23},
  pages = {034052},
  year = {2025},
  doi = {10.1103/PhysRevApplied.23.034052}
}

@article{Stas2022,
  author = {Stas, P.-J. and Huan, Y. Q. and Machielse, B. and Knall, E. N. and Suleymanzade, A. and Pingault, B. and Sutula, M. and Ding, S. W. and Knaut, C. M. and Assumpcao, D. R. and Wei, Y.-C. and Bhaskar, M. K. and Riedinger, R. and Sukachev, D. D. and Park, H. and Loncar, M. and Levonian, D. S. and Lukin, M. D.},
  title = {Robust multi-qubit quantum network node with integrated error detection},
  journal = {Science},
  volume = {378},
  number = {6619},
  pages = {557--560},
  year = {2022},
  doi = {10.1126/science.add9771}
}

@article{Knaut2024,
  author = {Knaut, C. M. and Suleymanzade, A. and Wei, Y.-C. and Assumpcao, D. R. and Stas, P.-J. and Huan, Y. Q. and Machielse, B. and Knall, E. N. and Sutula, M. and Baranes, G. and Sinclair, N. and De-Eknamkul, C. and Levonian, D. S. and Bhaskar, M. K. and Park, H. and Loncar, M. and Lukin, M. D.},
  title = {Entanglement of nanophotonic quantum memory nodes in a telecom network},
  journal = {Nature},
  volume = {629},
  pages = {573--578},
  year = {2024},
  doi = {10.1038/s41586-024-07252-z}
}

@article{Ademi2025,
  author = {Zarije Ademi and Marion Bassi and C{\'e}cile X. Yu and Sander L. de Snoo and Stefan D. Oosterhout and Amir Sammak and Lieven M. K. Vandersypen and Giordano Scappucci and Corentin D{\'e}pr{\'e}z and Menno Veldhorst},
  title = {Distributing entanglement between distant semiconductor qubit registers using a shared-control shuttling link},
  journal = {arXiv:2510.26860},
  year = {2025}
}

@article{Steinacker2025,
  author = {P. Steinacker and N. Dumoulin Stuyck and W. H. Lim and others},
  title = {Industry-compatible silicon spin-qubit unit cells exceeding 99\% fidelity},
  journal = {Nature},
  volume = {646},
  pages = {81--87},
  year = {2025},
  doi = {10.1038/s41586-025-09531-9}
}

@article{Evered2023,
  author = {Evered, Simon J. and Bluvstein, Dolev and Kalinowski, Marcin and Ebadi, Sepehr and Manovitz, Tom and Zhou, Hengyun and Li, Sophie H. and Geim, Alexandra A. and Wang, Tout T. and Maskara, Nishad and Lukin, Mikhail D.},
  title = {High-fidelity parallel entangling gates on a neutral-atom quantum computer},
  journal = {Nature},
  volume = {622},
  pages = {268--272},
  year = {2023},
  doi = {10.1038/s41586-023-06481-y}
}

@article{vanLeent2022,
  author = {van Leent, Tim and Bock, Matthias and Fertig, Florian and Garthoff, Robert and Eppelt, Sebastian and Zhou, Yiru and Malik, Pooja and Seubert, Matthias and Bauer, Tobias and Rosenfeld, Wenjamin and Zhang, Wei and Becher, Christoph and Weinfurter, Harald},
  title = {Entangling single atoms over 33 km telecom fibre},
  journal = {Nature},
  volume = {607},
  pages = {69--73},
  year = {2022},
  doi = {10.1038/s41586-022-04764-4}
}

@article{Li2024,
  author = {Li, Rui and Kubo, Kentaro and Ho, Yinghao and Yan, Zhiguang and Nakamura, Yasunobu and Goto, Hayato},
  title = {Realization of High-Fidelity CZ Gate Based on a Double-Transmon Coupler},
  journal = {Phys. Rev. X},
  volume = {14},
  pages = {041050},
  year = {2024},
  doi = {10.1103/PhysRevX.14.041050}
}

@article{Leung2019,
  author = {Leung, N. and Lu, Y. and Chakram, S. and Naik, R. K. and Earnest, N. and Ma, R. and Jacobs, K. and Cleland, A. N. and Schuster, D. I.},
  title = {Deterministic bidirectional communication and remote entanglement generation between superconducting qubits},
  journal = {npj Quantum Information}, volume = {5}, pages = {18}, year = {2019}, doi = {10.1038/s41534-019-0128-0}
}

@article{Weber2024,
  author = {Weber, M. A. and Gely, M. F. and Hanley, R. K. and Harty, T. P. and Leu, A. D. and L\"oschnauer, C. M. and Nadlinger, D. P. and Lucas, D. M.},
  title = {Robust and fast microwave-driven quantum logic for trapped-ion qubits},
  journal = {Physical Review A}, volume = {110}, pages = {L010601}, year = {2024},
  doi = {10.1103/PhysRevA.110.L010601}
}

@article{Stephenson2020,
  author = {Stephenson, L. J. and Nadlinger, D. P. and Nichol, B. C. and An, S. and Drmota, P. and Ballance, T. G. and Thirumalai, K. and Goodwin, J. F. and Lucas, D. M. and Ballance, C. J.},
  title = {High-Rate, High-Fidelity Entanglement of Qubits Across an Elementary Quantum Network},
  journal = {Physical Review Letters}, volume = {124}, pages = {110501}, year = {2020},
  doi = {10.1103/PhysRevLett.124.110501}
}

@article{Pingault2017,
author = {Pingault, Benjamin and Jarausch, David-Dominik and Hepp, Christian and Lina, Eti and Becker, Jonas and Markham, Matthew and Becher, Christoph and Atatüre, Mete},
year = {2017},
month = {01},
pages = {},
title = {Coherent control of the silicon-vacancy spin in diamond},
volume = {8},
journal = {Nature Communications},
doi = {10.1038/ncomms15579}
}

@article{Magnard2020,
  title = {Microwave Quantum Link between Superconducting Circuits Housed in Spatially Separated Cryogenic Systems},
  author = {Magnard, P. and Storz, S. and Kurpiers, P. and Sch{\"a}r, J. and Marxer, F. and L{\"u}tolf, J. and Walter, T. and Besse, J.-C. and Gabureac, M. and Reuer, K. and Akin, A. and Royer, B. and Blais, A. and Wallraff, A.},
  journal = {Phys. Rev. Lett.},
  volume = {125},
  issue = {26},
  pages = {260502},
  numpages = {7},
  year = {2020},
  month = {Dec},
  publisher = {American Physical Society},
  doi = {10.1103/PhysRevLett.125.260502},
  url = {https://link.aps.org/doi/10.1103/PhysRevLett.125.260502}
}

@article{Giedke1999,
  title   = {Lower bounds for attainable fidelity in entanglement purification},
  author  = {Giedke, G. and Briegel, H.-J. and Cirac, J. I. and Zoller, P.},
  journal = {Phys. Rev. A}, volume = {59}, pages = {2641}, year = {1999},
  doi     = {10.1103/PhysRevA.59.2641}
}

@article{DurBriegel2007,
  title   = {Entanglement purification and quantum error correction},
  author  = {D{\"u}r, W. and Briegel, H.-J.},
  journal = {Rep. Prog. Phys.}, volume = {70}, pages = {1381}, year = {2007}
}

@misc{Rohde2007,
  title   = {Noise thresholds for entanglement purification},
  author  = {Rohde, Peter P.},
  year    = {2007}, eprint = {quant-ph/0702065}, archivePrefix = {arXiv}
}

@article{FujiiYamamoto2009,
  title   = {Entanglement purification with double selection},
  author  = {Fujii, Keisuke and Yamamoto, Katsuji},
  journal = {Phys. Rev. A}, volume = {80}, pages = {042308}, year = {2009},
  doi     = {10.1103/PhysRevA.80.042308}
}

@article{Zwerger2013,
  title   = {Universal and Optimal Error Thresholds for Measurement-Based Entanglement Purification},
  author  = {Zwerger, M. and Briegel, H. J. and D{\"u}r, W.},
  journal = {Phys. Rev. Lett.}, volume = {110}, pages = {260503}, year = {2013},
  doi     = {10.1103/PhysRevLett.110.260503}
}

@article{Krastanov2019,
  title   = {Optimized Entanglement Purification},
  author  = {Krastanov, Stefan and Albert, Victor V. and Jiang, Liang},
  journal = {Quantum}, volume = {3}, pages = {123}, year = {2019},
  doi     = {10.22331/q-2019-02-18-123}
}

@article{Jansen2022,
  title   = {Enumerating all bilocal {C}lifford distillation protocols through symmetry reduction},
  author  = {Jansen, Sarah and Goodenough, Kenneth and de Bone, S{\'e}bastian and Gijswijt, Dion and Elkouss, David},
  journal = {Quantum}, volume = {6}, pages = {715}, year = {2022},
  doi     = {10.22331/q-2022-05-19-715}
}

@article{PretiBernad2024,
  title   = {Statistical evaluation and optimization of entanglement purification protocols},
  author  = {Preti, Francesco and Bern{\'a}d, J{\'o}zsef Zsolt},
  journal = {Phys. Rev. A}, volume = {110}, pages = {022619}, year = {2024},
  doi     = {10.1103/PhysRevA.110.022619}
}

@article{Siddhu2025,
  title   = {Basic entanglement distillation with realistic noise},
  author  = {Siddhu, Vikesh and Winston, Erick and McKay, David C. and Javadi-Abhari, Ali},
  journal = {Phys. Rev. Research}, volume = {7}, pages = {043069}, year = {2025},
  doi     = {10.1103/xx9g-pkht}
}

@article{KimYunBae2025,
  title   = {Purification of noisy measurements and faithful distillation of entanglement},
  author  = {Kim, Jaemin and Yun, Jiyoung and Bae, Joonwoo},
  journal = {J. Phys. A: Math. Theor.}, year = {2025},
  doi     = {10.1088/1751-8121/ada0fa}
}

@article{Kim2026,
  title   = {Protocol for purifying noisy preparation and measurements of qubits},
  author  = {Kim, J. and Seo, S. and Yun, J. and Lienhard, B. and Bae, J.},
  journal = {Phys. Rev. Research}, volume = {8}, pages = {023347}, year = {2026},
  doi     = {10.1103/z6pg-m3q6}
}

@article{AuerBurkard2014,
  title   = {Entanglement purification with the exchange interaction},
  author  = {Auer, Adrian and Burkard, Guido},
  journal = {Phys. Rev. A}, volume = {90}, pages = {022320}, year = {2014},
  doi     = {10.1103/PhysRevA.90.022320}
}

@article{Zang2025,
  title   = {No-Go Theorems for Universal Entanglement Purification},
  author  = {Zang, Allen and Chen, Xinan and Chitambar, Eric and Suchara, Martin and Zhong, Tian},
  journal = {Phys. Rev. Lett.}, volume = {134}, pages = {190803}, year = {2025},
  doi     = {10.1103/PhysRevLett.134.190803}
}

@misc{Tirone2026,
  title   = {Universal Entanglement Distillation},
  author  = {Tirone, Salvatore and Mele, Francesco Anna and Giovannetti, Vittorio and Lami, Ludovico},
  year    = {2026}, eprint = {2609.40080}, archivePrefix = {arXiv}
}

@article{Pan2003,
  title   = {Experimental entanglement purification of arbitrary unknown states},
  author  = {Pan, Jian-Wei and Gasparoni, Sara and Ursin, Rupert and Weihs, Gregor and Zeilinger, Anton},
  journal = {Nature}, volume = {423}, pages = {417}, year = {2003},
  doi     = {10.1038/nature01623}
}

@article{Reichle2006,
  title   = {Experimental purification of two-atom entanglement},
  author  = {Reichle, R. and others},
  journal = {Nature}, volume = {443}, pages = {838}, year = {2006},
  doi     = {10.1038/nature05146}
}

@article{Kalb2017,
  title   = {Entanglement distillation between solid-state quantum network nodes},
  author  = {Kalb, N. and others},
  journal = {Science}, volume = {356}, pages = {928}, year = {2017},
  doi     = {10.1126/science.aan0070}
}

@article{Hu2021,
  title   = {Long-Distance Entanglement Purification for Quantum Communication},
  author  = {Hu, Xiao-Min and Huang, Cen-Xiao and Sheng, Yu-Bo and Zhou, Lan and Liu, Bi-Heng and Guo, Yu and Zhang, Chao and Xing, Wen-Bo and Huang, Yun-Feng and Li, Chuan-Feng and Guo, Guang-Can},
  journal = {Phys. Rev. Lett.}, volume = {126}, pages = {010503}, year = {2021},
  doi     = {10.1103/PhysRevLett.126.010503}
}

@article{Ecker2021,
  title   = {Experimental Single-Copy Entanglement Distillation},
  author  = {Ecker, Sebastian and Sohr, Philipp and Bulla, Lukas and Huber, Marcus and Bohmann, Martin and Ursin, Rupert},
  journal = {Phys. Rev. Lett.}, volume = {127}, pages = {040506}, year = {2021},
  doi     = {10.1103/PhysRevLett.127.040506}
}

@article{Yan2022,
  title   = {Entanglement Purification and Protection in a Superconducting Quantum Network},
  author  = {Yan, Haoxiong and Zhong, Youpeng and Chang, Hung-Shen and Bienfait, Audrey and Chou, Ming-Han and Conner, Christopher R. and Dumur, {\'E}tienne and Grebel, Joel and Povey, Rhys G. and Cleland, Andrew N.},
  journal = {Phys. Rev. Lett.}, volume = {128}, pages = {080504}, year = {2022},
  doi     = {10.1103/PhysRevLett.128.080504}
}

@unpublished{Alessio_2026,
  author = {Alessio Belshi and Andr{\'a}s Kiny{\'o} and Alessia Camuti Borani and Tommaso Calarco and Felix Motzoi and Zolt\''{a}n Zimbor\'{a}s and Orsolya K{\'a}lm{\'a}n and Matthias M. M\"{u}ller},
  title  = {Quantifying  Entanglement Distillation for Interconnected Color Centers},
  note   = {In preparation}
}

\appendix

\section{Estimating effective error parameters}
\label{app:A} 
In what follows we present how we convert the measurable physical properties of quantum platforms to error parameters that we can use in our models. We note that these estimates serve only as guidelines -- providing the correct orders of magnitude -- while there exist  more sophisticated methods for correlating such quantities.

\subsection{Preparation noise}

In experimental papers the preparation fidelity is the quantity that is often published. Since we know that the fidelity of a state with white noise parameter $\varepsilon$ to the pure Bell-state is
\[
\mathcal F(\varepsilon) = 1-\dfrac{3\varepsilon}{4},
\]
an effective parameter can be associated to a state with a given fidelity, by assuming that all infidelities can be attributed to white noise:
\[ \varepsilon_{\rm eff.} = \dfrac{4}{3}(1-\mathcal F). \]

\subsection{Depolarizing channel}
The depolarizing channel's parameter is most straightforwardly computed from the average gate infidelity, $r$. Since we are working with a two-qubit depolarization channel, the conversion is:
\[ \lambda = \dfrac{4}{3}r, \]
similarly to the case of preparation noise.

\section{Reduction of asymmetric gate errors to symmetric ones}
\label{app:B}
In the following we show that the output fidelity depends on $\lambda_A+\lambda_B$ in first order, which provides a way to reduce asymmetric gate errors to symmetric ones. This is most conveniently done by introducing the map which transforms the states:
\begin{equation}
    \rho_{n+1}(\lambda_A,\lambda_B) = \mathcal{P}_{\lambda_A,\lambda_B}[\rho_n(\lambda_A,\lambda_B)],
\end{equation}
with
\begin{equation}
    \mathcal{P}_{\lambda_A,\lambda_B}(\rho) = \dfrac{\Tr_{2,4}(P_0 (\mathcal E^{\lambda_A}\otimes \mathcal E^{\lambda_B})[\Omega'] P_0)}{\Tr(P_0 (\mathcal E^{\lambda_A}\otimes \mathcal E^{\lambda_B})[\Omega'] P_0)},
\end{equation}
and $\Omega' = (U\otimes U)S_{23}(\rho\otimes\rho)S_{23}(U^{\dagger}\otimes U^{\dagger})$.

Since the whole setup is symmetric to exchanging Alice and Bob, the following identity holds:
\begin{equation}
    \label{eq:B-proof-symm}
    \mathcal P_{\lambda_A,\lambda_B}(\rho) = S \cdot \mathcal P_{\lambda_B,\lambda_A}(S\rho S) \cdot S,
\end{equation}
where $S$ denotes the swap operation on the two qubits constituting the state $\rho$ (for both copies of the input), as well as in the state after post-selection. 

Now we show by induction that for arbitrary $n$ the output state after $n$ iterations obeys $S\cdot \rho_n(\lambda_A,\lambda_B)\cdot S = \rho_n(\lambda_B, \lambda_A)$, given that the input state is swap-invariant, $S\rho_0 S=\rho_0$ (which holds in our case). After the first iteration, from \eqref{eq:B-proof-symm} we get
\begin{equation}
    \rho_1(\lambda_A, \lambda_B) = \mathcal P_{\lambda_A, \lambda_B}(\rho_0) = S \cdot \mathcal P_{\lambda_B,\lambda_A}(\rho_0) \cdot S,
\end{equation}
thus $S\cdot \rho_1(\lambda_A, \lambda_B) \cdot S = \rho_1(\lambda_B, \lambda_A)$.
Let us assume that $S\cdot \rho_n(\lambda_A,\lambda_B)\cdot S = \rho_n(\lambda_B, \lambda_A)$, then after one more iteration:
\begin{equation}
    \rho_{n+1}(\lambda_A, \lambda_B) = \mathcal P_{\lambda_A, \lambda_B}[\rho_n(\lambda_A,\lambda_B)].
\end{equation}
After swapping the qubits, using \eqref{eq:B-proof-symm} we get
\begin{equation}
    S\cdot \rho_{n+1}(\lambda_A, \lambda_B)\cdot S = \mathcal{P}_{\lambda_B,\lambda_A}[S\cdot \rho_n(\lambda_A,\lambda_B)\cdot S].
\end{equation}
Using the induction hypothesis we can rewrite this as
\begin{equation}
    \mathcal P_{\lambda_B, \lambda_A}[\rho_n(\lambda_B, \lambda_A)] = \rho_{n+1}(\lambda_B, \lambda_A).
\end{equation}

Let us now take a swap-invariant hermitian operator $[O,S]=0$. Then the expectation value of $O$ in the state $\rho_n(\lambda_A, \lambda_B)$ is symmetric for the $\lambda_A\leftrightarrow\lambda_B$ exchange:
\begin{equation}
    \expval{O}_{\lambda_A, \lambda_B, n} = \Tr(O\cdot \rho_{n}(\lambda_A,\lambda_B)) = \Tr(O\cdot \rho_{n}(\lambda_B,\lambda_A)),
\end{equation}
for arbitrary $n$. Then the derivative of this with respect to $\lambda_A$ and $\lambda_B$ will equal 
\begin{equation}
    \partial_{\lambda_A} \expval{O}|_{(0,0)} = \partial_{\lambda_B} \expval{O}|_{(0,0)},
\end{equation}
where, for simplicity, we used the notation $\expval{O}_{\lambda_A, \lambda_B, n} = \expval{O}$.

Taking the Taylor expansion of the expectation of $O$ gives
\begin{multline}
    \expval{O}_{\lambda_A, \lambda_B, n} =\expval{O}_{0, 0, n} + \partial_{\lambda_A} \expval{O}|_{(0,0)} \cdot \lambda_A +\\
    +\partial_{\lambda_B} \expval{O}|_{(0,0)}\cdot \lambda_B + \order{\lambda^2} = \\
    = \expval{O}_{0, 0, n} + \partial_{\lambda_A} \expval{O}|_{(0,0)} \cdot (\lambda_A + \lambda_B)+\order{\lambda^2}.
\end{multline}

Choosing $O = \ket{\Phi_+}\bra{\Phi_+}$, the expectation value becomes the output fidelity and one can readily see that it depends on $\lambda_{A}+\lambda_{B}$ in first order.

For coherent errors, $\mathcal E^{\theta}$ can be chosen as
\begin{equation}
    \mathcal E^{\theta}(\rho) = e^{-i\theta H}\rho e^{i\theta H},
\end{equation}
and the proof holds there as well. We note however, that the types of coherent errors used in this paper generate a fidelity which is even in the error parameter: $\mathcal F(\theta) = \mathcal F(-\theta)$, thus its first derivative vanishes, and therefore has no first-order contribution. For readout errors, \eqref{eq:B-proof-symm} can be similarly proven, and thus the proof holds as well.

\section{Success probability enhancement - proofs}
\label{app:C}

In this Appendix we derive the conditions under which post-selection on the measurement outcome '$00$' leads to the same distillation dynamics as post-selection on '$11$'. We first consider the ideal protocol and determine the corresponding condition on the input state. We then extend the analysis to noisy local operations.

Throughout this Appendix, we denote the set of single-qubit Pauli operators by
\begin{equation}
    \mathcal{P}=\{I,X,Y,Z\}.
\end{equation}
Rather than assigning a single numerical index to a two-qubit Pauli string, we label it explicitly by the pair of Pauli operators acting on the two qubits.

\subsection{Conditions for the input state}

An arbitrary two-qubit state can be expanded as
\begin{equation}
    \rho = \frac{1}{4}\sum_{P,Q\in\mathcal{P}} r_{PQ}\,P\otimes Q,
    \qquad
    r_{PQ}=\Tr\!\left[(P\otimes Q)\rho\right].
    \label{eq:appC_pauli_expansion}
\end{equation}
Since $\rho$ is Hermitian, all coefficients $r_{PQ}$ are real, and $r_{II}=1$ for a normalized state.

The input to one iteration consists of two identical copies of $\rho$. After exchanging qubits 2 and 3, the four-qubit state is
\begin{equation}
\begin{aligned}
    S_{23}(\rho\otimes\rho)S_{23}^{\dagger}
    = \frac{1}{16}
    \sum_{P,Q,R,S\in\mathcal{P}}
    r_{PQ}r_{RS}\,
    (P\otimes R)\otimes(Q\otimes S).
\end{aligned}
\label{eq:appC_after_swap}
\end{equation}
The first two-qubit factor belongs to Alice and the second to Bob.

Both parties subsequently apply
\begin{equation}
    U=(H\otimes I)U_{\rm cnot}(I\otimes X).
    \label{eq:appC_U}
\end{equation}
Since $U$ is a Clifford operation, conjugation by $U$ maps every two-qubit Pauli string to another Pauli string up to a sign. For the unitary in Eq.~\eqref{eq:appC_U},
\begin{equation}
\begin{array}{c|cccc}
P\backslash Q & I & X & Y & Z \\ \hline
I & II & IX & -XY & -XZ \\
X & ZX & ZI & YZ & -YY \\
Y & -YX & -YI & ZZ & -ZY \\
Z & XI & XX & -IY & -IZ
\end{array}
\label{eq:appC_clifford_table}
\end{equation}
where, for example, $ZX$ denotes $Z\otimes X$. We write this action compactly as
\begin{equation}
    U(P\otimes Q)U^{\dagger}
    =\chi_{PQ}\,A_{PQ}\otimes B_{PQ},
    \qquad \chi_{PQ}\in\{+1,-1\},
    \label{eq:appC_clifford_action}
\end{equation}
with $A_{PQ},B_{PQ}\in\mathcal P$.

The measurement of the second qubit on each side can be incorporated through
\begin{equation}
    m_i(P)=\bra{i}P\ket{i},\qquad i\in\{0,1\},
\end{equation}
for which
\begin{equation}
    m_i(I)=1,\qquad
    m_i(Z)=(-1)^i,\qquad
    m_i(X)=m_i(Y)=0.
    \label{eq:appC_measurement_paulis}
\end{equation}
Consequently,
\begin{equation}
\begin{aligned}
    &(I\otimes\bra{i})
    U(P\otimes Q)U^{\dagger}
    (I\otimes\ket{i})
    \\
    &\hspace{2cm}
    =\chi_{PQ}\,m_i(B_{PQ})A_{PQ}.
\end{aligned}
\label{eq:appC_local_postselection}
\end{equation}
Only terms whose Pauli operator on the measured qubit is $I$ or $Z$ survive. The $I$ terms contribute with the same sign for the two outcomes, while the $Z$ terms change sign.

Let $\rho_i'$ denote the unnormalized state of the two remaining qubits when the outcome '$ii$' is accepted. Equations~\eqref{eq:appC_after_swap} and \eqref{eq:appC_local_postselection} give
\begin{equation}
\begin{aligned}
    \rho_i'
    =\frac{1}{16}
    \sum_{P,Q,R,S\in\mathcal P}
    &r_{PQ}r_{RS}\,
    \chi_{PR}\chi_{QS}\,
    m_i(B_{PR})m_i(B_{QS})
    \\
    &\times A_{PR}\otimes A_{QS}.
\end{aligned}
\label{eq:appC_output_general}
\end{equation}

We first require
\begin{equation}
    \rho_0'=\rho_1'.
    \label{eq:appC_first_condition}
\end{equation}
The calculation is considerably more transparent if we inspect the Pauli coefficients of the difference
$\Delta\equiv\rho_0'-\rho_1'$. In particular, Eq.~\eqref{eq:appC_output_general} gives
\begin{align}
    \Tr(II\,\Delta)
        = -\frac{1}{2}&\left(r_{IZ}^{2}+r_{ZI}^{2}\right),
        \label{eq:appC_delta_II}\\
    \Tr(ZZ\,\Delta)
        = \frac{1}{2}&\left(r_{XY}^{2}+r_{YX}^{2}\right),
        \label{eq:appC_delta_ZZ}\\
    \Tr(IZ\,\Delta)
        = \frac{1}{2}\left(r_{IY}^{2}-r_{ZX}^{2}\right),
        &\quad
    \Tr(XY\,\Delta)=r_{IY}r_{ZX},
        \label{eq:appC_delta_IY_ZX}\\
    \Tr(ZI\,\Delta)
        = \frac{1}{2}\left(r_{YI}^{2}-r_{XZ}^{2}\right),
        &\quad
    \Tr(YX\,\Delta)=r_{XZ}r_{YI}.
        \label{eq:appC_delta_XZ_YI}
\end{align}
Because the $r_{PQ}$ are real, Eqs.~\eqref{eq:appC_delta_II} and \eqref{eq:appC_delta_ZZ} imply
\begin{equation}
    r_{IZ}=r_{ZI}=r_{XY}=r_{YX}=0.
\end{equation}
Moreover, the simultaneous equations $a^2-b^2=0$ and $ab=0$ have only the real solution $a=b=0$. Hence Eqs.~\eqref{eq:appC_delta_IY_ZX} and \eqref{eq:appC_delta_XZ_YI} further imply
\begin{equation}
    r_{IY}=r_{ZX}=r_{XZ}=r_{YI}=0.
\end{equation}
Conversely, direct substitution into Eq.~\eqref{eq:appC_output_general} shows that these eight conditions make every Pauli coefficient of $\Delta$ vanish. Therefore there are no additional nonlinear branches of solutions, and
\begin{equation}
\begin{aligned}
    \rho_0'=\rho_1'
    \quad\Longleftrightarrow\quad
    &r_{IY}=r_{IZ}=r_{XY}=r_{XZ}\\
    &=r_{YI}=r_{YX}=r_{ZI}=r_{ZX}=0.
\end{aligned}
\label{eq:appC_first_coefficients}
\end{equation}
These conditions are precisely equivalent to
\begin{equation}
    [\rho,X\otimes X]=0.
    \label{eq:appC_commute_xx}
\end{equation}

This condition guarantees equality of the '$00$' and '$11$' outputs for one iteration. For the two post-selection schemes to generate the same \emph{dynamics}, the common output must itself satisfy the same symmetry. We therefore additionally require
\begin{equation}
    [\rho_i',X\otimes X]=0.
    \label{eq:appC_closure}
\end{equation}
After imposing Eq.~\eqref{eq:appC_first_coefficients}, the only potentially nonzero Pauli coefficients of this commutator are proportional to
\begin{equation}
    r_{IX}r_{ZY},\qquad
    r_{IX}^{2}-r_{ZY}^{2},\qquad
    r_{XI}r_{YZ},\qquad
    r_{XI}^{2}-r_{YZ}^{2}.
    \label{eq:appC_closure_polynomials}
\end{equation}
The same real-variable argument used above therefore shows that Eq.~\eqref{eq:appC_closure} holds if and only if
\begin{equation}
    r_{IX}=r_{XI}=r_{YZ}=r_{ZY}=0.
    \label{eq:appC_second_coefficients}
\end{equation}
Equivalently,
\begin{equation}
    [\rho,Z\otimes Z]=0.
    \label{eq:appC_commute_zz}
\end{equation}

Combining Eqs.~\eqref{eq:appC_commute_xx} and \eqref{eq:appC_commute_zz}, the required condition is
\begin{equation}
    [\rho,X\otimes X]=[\rho,Z\otimes Z]=0.
    \label{eq:appC_combined_condition}
\end{equation}
The operators $X\otimes X$ and $Z\otimes Z$ commute, and their common eigenvectors are precisely the four Bell states. Hence Eq.~\eqref{eq:appC_combined_condition} holds if and only if $\rho$ is Bell diagonal. Such a state can be written as
\begin{equation}
    \rho_{\rm BD}
    =\frac{1}{4}\left(
        I\otimes I+c_xX\otimes X+c_yY\otimes Y+c_zZ\otimes Z
    \right),
    \label{eq:appC_bell_diagonal}
\end{equation}
with real coefficients $c_x,c_y,c_z$ restricted by positivity. Thus, in the absence of gate errors, Bell-diagonal states form exactly the class of input states for which accepting '$00$' and accepting '$11$' generate identical distillation dynamics.

\subsection{Conditions for the error channel \label{app:C_2}}

We now include a local two-qubit error channel $\mathcal E$ after the application of $U$. We assume that the same channel acts on Alice's and Bob's subsystems. The action of $\mathcal E$ in the Pauli basis is described by its Pauli transfer matrix (PTM):
\begin{equation}
    \mathcal E(P\otimes Q)
    =\sum_{A,B\in\mathcal P}R_{PQ;AB}\,A\otimes B,
    \label{eq:appC_ptm}
\end{equation}
where
\begin{equation}
    R_{PQ;AB}
    =\frac{1}{4}\Tr\!\left[(A\otimes B)\,\mathcal E(P\otimes Q)\right].
    \label{eq:appC_ptm_elements}
\end{equation}

For a Bell-diagonal input it is convenient to write Eq.~\eqref{eq:appC_bell_diagonal} as
\begin{equation}
    \rho_{\rm BD}=\frac{1}{4}\sum_{P\in\mathcal P}c_P\,P\otimes P,
    \qquad c_I=1.
    \label{eq:appC_bell_compact}
\end{equation}
After exchanging qubits 2 and 3,
\begin{equation}
    S_{23}(\rho_{\rm BD}\otimes\rho_{\rm BD})S_{23}^{\dagger}
    =\frac{1}{16}\sum_{P,Q\in\mathcal P}c_Pc_Q\,
    (P\otimes Q)\otimes(P\otimes Q).
    \label{eq:appC_bell_after_swap}
\end{equation}
Let $\Gamma_{PQ}$ denote the Pauli string obtained from $P\otimes Q$ under conjugation by $U$, with its overall sign omitted:
\begin{equation}
    U(P\otimes Q)U^{\dagger}=\chi_{PQ}\Gamma_{PQ}.
\end{equation}
Because the same string occurs on Alice's and Bob's sides, the sign appears twice and cancels. Hence, after the local unitaries,
\begin{equation}
    \Xi
    =\frac{1}{16}\sum_{P,Q\in\mathcal P}c_Pc_Q\,
    \Gamma_{PQ}\otimes\Gamma_{PQ}.
    \label{eq:appC_Xi_bell}
\end{equation}

To combine the noise and post-selection into a compact notation, define
\begin{equation}
    v_i^{PQ}(A)
    =R_{\Gamma_{PQ};AI}+(-1)^iR_{\Gamma_{PQ};AZ},
    \qquad A\in\mathcal P.
    \label{eq:appC_v_definition}
\end{equation}
Indeed,
\begin{equation}
\begin{aligned}
    &(I\otimes\bra{i})\mathcal E(\Gamma_{PQ})(I\otimes\ket{i})\\
    &\hspace{1cm}=\sum_{A\in\mathcal P}v_i^{PQ}(A)A.
\end{aligned}
\end{equation}
The unnormalized post-selected state is therefore
\begin{equation}
    \rho_i'
    =\frac{1}{16}\sum_{P,Q\in\mathcal P}c_Pc_Q
    \sum_{A,C\in\mathcal P}
    v_i^{PQ}(A)v_i^{PQ}(C)\,A\otimes C.
    \label{eq:appC_noisy_output}
\end{equation}
This expression makes the two requirements on the error channel particularly transparent.

\begin{figure}[h]
\centering
\includegraphics[width=\linewidth]{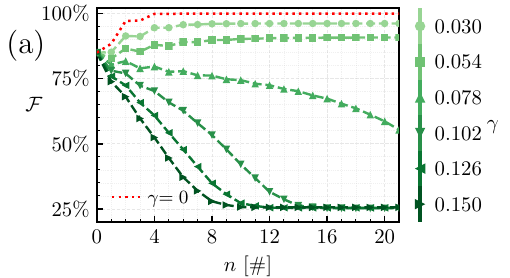}\\
\includegraphics[width=\linewidth]{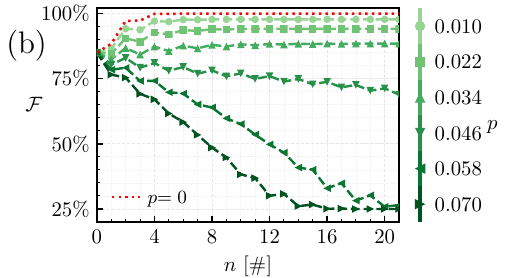}\\
\includegraphics[width=\linewidth]{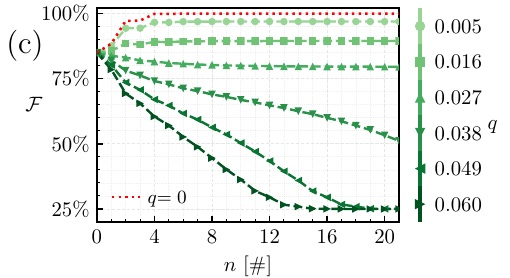}
\caption{Fidelity of the output state with respect to $\ket{\Phi_+}$ after successive protocol iterations. The states were initialized as $\rho_0(\varepsilon=0.2)$, and each curve corresponds to a different value of (a) amplitude damping parameter $\gamma$; (b) bit flip parameter $p$; (c) phase flip parameter $q$.}
\label{fig:16_other_steps}
\end{figure}

\subsubsection{Preservation of Bell diagonality}

First, a Bell-diagonal input must remain Bell diagonal after one post-selected iteration. Thus all coefficients of $A\otimes C$ with $A\neq C$ in Eq.~\eqref{eq:appC_noisy_output} must vanish. Requiring this for arbitrary Bell-diagonal input gives
\begin{equation}
    v_i^{PP}(A)v_i^{PP}(C)=0,
    \qquad (A\neq C),
    \label{eq:appC_BD_diag_condition}
\end{equation}
for $P\in\mathcal P$, together with
\begin{equation}
\begin{aligned}
    &v_i^{PQ}(A)v_i^{PQ}(C)
    +v_i^{QP}(A)v_i^{QP}(C)=0,\,\, (P\neq Q,\, A\neq C).
\end{aligned}
\label{eq:appC_BD_cross_condition}
\end{equation}
These conditions must hold for $i=0,1$. They are simply the PTM conditions stating that the post-selected map does not generate components outside the Bell-diagonal Pauli sector
\begin{equation}
    \mathcal B=\operatorname{span}\{II,XX,YY,ZZ\}.
\end{equation}

\subsubsection{Equality of the '00' and '11' outputs}

The second requirement is
\begin{equation}
    \rho_0'=\rho_1'.
    \label{eq:appC_noisy_symmetry}
\end{equation}
Assuming the Bell-diagonality conditions above, it is sufficient to compare the coefficients of $A\otimes A$. From Eq.~\eqref{eq:appC_v_definition},
\begin{equation}
\begin{aligned}
    [v_0^{PQ}(A)]^2-[v_1^{PQ}(A)]^2
    =4R_{\Gamma_{PQ};AI}R_{\Gamma_{PQ};AZ}.
\end{aligned}
\label{eq:appC_v_difference}
\end{equation}
Defining
\begin{equation}
    D_{\Gamma}(A)=R_{\Gamma;AI}R_{\Gamma;AZ},
\end{equation}
condition~\eqref{eq:appC_noisy_symmetry} for arbitrary Bell-diagonal inputs becomes
\begin{equation}
    D_{\Gamma_{PP}}(A)=0,
    \qquad P\in\mathcal P,
    \label{eq:appC_D_diag}
\end{equation}
and
\begin{equation}
    D_{\Gamma_{PQ}}(A)+D_{\Gamma_{QP}}(A)=0,
    \qquad P\neq Q,
    \label{eq:appC_D_cross}
\end{equation}
for every $A\in\mathcal P$.

Together, Eqs.~\eqref{eq:appC_BD_diag_condition}--\eqref{eq:appC_D_cross} give the conditions on the PTM under which the Bell-diagonal family is preserved and the '$00$' and '$11$' post-selections produce the same unnormalized output state. Although these conditions are quadratic in the PTM elements, they are straightforward to verify for a specified channel. In particular, any two-qubit channel that has a diagonal PTM satisfies them. Consequently, the depolarizing, bit-flip, and phase-flip channels preserve both Bell diagonality and the $00$--$11$ symmetry. By contrast, amplitude damping and the coherent error models considered in Sec.~IV do not in general satisfy these conditions.

\section{Supplementary figures}
\label{app:D}

In this section we present figures showing how the output fidelity is changing with every iteration of the protocol if one includes amplitude damping, bit-flip, or phase flip errors in the gates (see Fig.~\ref{fig:16_other_steps}(a)-(c)).
\end{document}